\documentclass[11pt, a4paper, onecolumn]{article}
\usepackage{bm}
\usepackage{graphicx}
\usepackage{subfigure}
\usepackage{amssymb}
\usepackage{amsmath}
\usepackage{algorithm}
\usepackage{algpseudocode}
\usepackage{tikz}
\usepackage{booktabs}
\usepackage{tcolorbox}
\usepackage{makeidx}

\usepackage[left=1.6cm, right=1.6cm, top=1.6cm, bottom=2cm]{geometry}

\usepackage{natbib}

{\endtcolorbox}

\makeindex                  
\usepackage{float}

\def\col{\textnormal{col}}

\newtheorem{definition}{Definition}
\newtheorem{assumption}{Assumption}

\newtheorem{theorem}{Theorem}

\def\ba{\begin{aligned}}
\def\ea{\end{aligned}}

\newcommand{\mV}{\mathbb{V}}
\newcommand{\mG}{\mathrm{G}}
\newcommand{\mE}{\mathbb{E}}

\newcommand{\R}{{\rm Re}}
\newcommand{\gc}{\mathrm{G}_{\rm d}}

\newcommand{\yb}{\mathbf{y}}

\newcommand{\Pb}{\mathsf{P}}

\newcommand{\Hs}{\mathsf{H}}

\newcommand{\Ps}{\mathsf{P}}

\newcommand{\Ls}{\mathsf{L}}

\newcommand{\Lc}{\mathcal{L}}
\newcommand{\Qs}{\mathsf{Q}}
\newcommand{\As}{\mathsf{A}}
\newcommand{\Bs}{\mathsf{B}}

\newcommand{\Is}{\mathsf{I}}
\newcommand{\Us}{\mathsf{U}}
\newcommand{\pb}{\mathsf{P}}

\newcommand{\sbf}{\mathbf{s}}

\newcommand{\eb}{\mathbf{e}}
\newcommand{\rb}{\mathbf{r}}

\newcommand{\rhb}{{\bm\rho}}

\newcommand{\xib}{\bm{\xi}}
\newcommand{\nub}{\tilde{\bm{\nu}}}

\title{\huge Blended Dynamics and Emergence\\ in Open Quantum Networks}

\author{%
    Qinghao Wen, Zihao Ren,
    Lei Wang,
    Hyungbo Shim, and
    Guodong Shi\thanks{Q. Wen, Z. Ren and G. Shi are with the Australian Centre for Robotics, The University of Sydney, NSW 2006, Australia (Email: qinghao.wen@uni.sydney.edu.au, zren0735@uni.sydney.edu.au, guodong.shi@sydney.edu.au). L. Wang is with College of Control Science and Engineering, Zhejiang University, Hangzhou, China (Email: lei.wangzju@zju.edu.cn). H. Shim is with Department of Electrical and Computer Engineering, Seoul National University, Korea (Email: hshim@snu.ac.kr). This work was supported by the Australian Research Council under Grant
DP190103615, Grant LP210200473,   Grant DP230101014,  and by the Faculty of Engineering's Breakthrough Project at The University of Sydney. The first two authors contributed equally to this work.}
}

\date{}
\begin{document}

\maketitle

\begin{abstract}
In this paper, we develop a blended dynamics framework for open quantum networks with diffusive couplings. The network consists of qubits interconnected through Hamiltonian couplings, environmental dissipation, and consensus-like diffusive interactions. Such networks commonly arise in spontaneous emission processes and non-Hermitian quantum computing, and their evolution follows a Lindblad master equation. Blended dynamics theory is well established in the classical setting as a tool for analyzing emergent behaviors in heterogeneous networks with diffusive couplings. Its key insight is to blend the local dynamics rather than the trajectories of individual nodes. Perturbation analysis then shows that, under sufficiently strong coupling, all node trajectories tend to stay close to those of the blended system over time. We first show that this theory extends naturally to the reduced-state dynamics of quantum networks, revealing classical-like clustering phenomena in which qubits converge to a shared equilibrium or a common trajectory determined by the quantum blended reduced-state dynamics. We then extend the analysis to qubit coherent states using quantum Laplacians and induced graphs, proving orbit attraction of the network density operator toward the quantum blended coherent dynamics,  establishing the emergence of intrinsically quantum and dynamically clustering behaviors. Finally, numerical examples validate the theoretical results.
\end{abstract}


\section{Introduction}

Quantum networks form the foundational infrastructure for quantum computing and quantum communication systems \cite{RevModPhys.82.1209,quantumcontrol1,quantumcontrol2,quantumcom1,quantumcom2}. Such networks consist of interconnected quantum nodes, such as photons, electrons, or atoms, whose interactions are mediated by non-classical correlations, including entanglement \cite{RevModPhys.82.1209}. In realistic quantum engineering settings, these networks inevitably interact with their surrounding environments, giving rise to dissipative effects through which quantum information is gradually leaked to the environment \cite{lind2}. Open quantum networks also arise naturally in a variety of physical scenarios, such as spontaneous emission in arrays of quantum emitters, which can lead to collective phenomena such as superradiance and subradiance \cite{emission}.  Engineered dissipation mechanisms have also been designed to steer the network toward strongly correlated or entangled steady states, as a principled way for realizing universal quantum computing \cite{nonH}. 

The dynamics of open quantum networks are commonly described by a quantum master equation first formalized in \cite{lind1}. These equations, now known as Lindblad master equations, have been extensively studied in the contexts of quantum computation, quantum information processing, and quantum control theory; see, for example, \cite{openquantumnet,quannet2}. A particularly interesting subclass is that of diffusively coupled open quantum networks, in which the interactions between nodes exhibit diffusive, consensus-like structures analogous to those studied in classical networked dynamical systems, e.g., \cite{swap_operator,Shi2016Reaching}. 

 In the discrete-time setting, \cite{swap_operator} first introduced quantum consensus based on swap operators and proposed a quantum gossip algorithm, establishing a framework encompassing four classes of consensus. Subsequent extensions addressed symmetrization within a group-theoretic context \cite{mazzarella2015}. Other notable works include \cite{Takeuchi2016Distributed}, which designed local observation and feedback strategies that preserve state purity, and \cite{Jafarizadeh2016Optimizing}, which connected convergence rates to the dimension of the local Hilbert space. Building on the theory of Lindblad master equations, \cite{Ticozzi2014Symmetrization,WCICA} proposed classes of master equations that induce consensus behaviors in open quantum networks, where convergence is established using Lyapunov-based analysis and invariance principles. Subsequently, \cite{Shi2016Reaching} established convergence of quantum network states to  consensus  by introducing the notions of  quantum Laplacian and induced graphs. Under this framework, quantum consensus was shown to be equivalent to component-wise consensus on the associated induced graphs.
 This idea was later extended to directed and switching networks \cite{Shi2017} as well as hybrid quantum networks \cite{Shi2017Reaching}. A comprehensive survey of these developments was provided in \cite{marcozzi2021quantumconsensusoverview}.

To date, only limited results have been reported on how more general collective behaviors such as clustering \cite{cluster} and synchronization \cite{synchron}, which are well established in classical networked systems, emerge in quantum networks. In \cite{Shi2016Reaching}, orbit synchronization was shown to be achievable for open quantum networks under strong symmetry conditions imposed on the system Hamiltonians. The development of systematic analytical tools for studying collective and genuinely non-classical emergent behaviors in quantum networks beyond consensus phenomena remains largely open.

Meanwhile, for classical networks with diffusive couplings, blended dynamics theory has been developed as a powerful framework for analyzing emergent behaviors. Blended dynamics, first introduced in  \cite{TAC2016HKim}, refers to a quasi-steady-state subsystem obtained by averaging the individual dynamics of agents that are strongly coupled. Then under sufficiently strong coupling, the blended dynamics accurately captures the collective behavior of multi-agent networks for even non-autonomous and nonlinear agent dynamics. The theory was subsequently extended to settings involving rank-deficient coupling structures in \cite{LEE2020108952}, and has since found applications in a wide range of multi-agent system problems, including distributed estimation and optimization; see, \cite{Lee2022}, for a comprehensive survey.

In this paper, we extend the classical blended dynamics framework to open quantum networks with diffusive couplings. We consider quantum networks in which nodes evolve under a network Hamiltonian,  subject to environmental dissipation and diffusive interactions. The blended dynamics for such quantum networks is first defined at the level of the reduced states of individual nodes, obtained by averaging the local node dynamics in a manner analogous to classical networked systems. When both the Hamiltonian and the dissipation operators are separable, the resulting {\em quantum blended reduced-state dynamics} recovers the classical blended dynamics theory, thereby characterizing the classical behaviors of the quantum network through quantum reduced-state averaging. Next, we exploit the induced graph associated with the quantum network and introduce the notion of {\em quantum blended coherent dynamics}. This concept is defined on subsystems that achieve consensus within each connected component of the induced graph, under which the emergence of genuinely quantum collective behaviors is rigorously established. The main results of this work are summarized as follows:
\begin{itemize}
    \item When the Hamiltonian couplings and environmental dissipation in diffusively coupled open quantum networks are separable, we show that the reduced states of the individual quantum nodes may converge either to a common equilibrium or evolve along a shared trajectory, both of which are determined by the blended reduced-state dynamics.
    \item For the general case in which the Hamiltonian couplings and environmental dissipation are inseparable, we construct a blended coherent dynamics and prove that, under sufficiently strong diffusive couplings, the network density operator {\em always} converges to an orbitally attractive, permutation-invariant subspace. 
    \item For each of the above scenarios, we rigorously characterize the consistency between the blended dynamics and the evolution of quantum network dynamics. In fact, we prove that the two remain arbitrarily close over any finite time interval, provided that the diffusive coupling gain is sufficiently large.   
\end{itemize}
It is worth emphasizing that, for each scenario, the coupling gain conditions required to guarantee such consistency are derived explicitly and constructively. These results offer new insights into the analysis and understanding of quantum network dynamics. To illustrate and validate the theoretical findings, representative examples are also presented.

The remainder of the paper is organized as follows. Section \ref{sec.pre} introduces the key concepts from quantum systems, graph theory, and classical blended dynamics. Section \ref{sec.dif} presents the diffusive and open quantum network models that form the basis of this work. In Section \ref{sec.cla}, we consider the case in which the Hamiltonian couplings and environmental dissipation are separable, establish the blended reduced-state dynamics, and analyze the classical behaviors of the quantum network. Section \ref{sec.eme} addresses the inseparable case, where we develop the blended coherent dynamics to investigate the coherent behaviors emerging in the network. Finally, Section \ref{sec.con} concludes the paper. All proofs are provided in the Appendices.

\section{Preliminaries}
\label{sec.pre}
In this section, we introduce key concepts from quantum systems in \cite{swap}, graph theory in \cite{mag} and classical blended dynamic system theory in \cite{Lee2022}. In addition, we define the notations used throughout this paper.

\subsection{Open Quantum Systems}
The state of an isolated qubit is fully described by a unit state vector \( |\psi\rangle \) defined in a two-dimensional Hilbert space \( \mathcal{H} \), which is a complex vector space equipped with an inner product. Based on this, the state space of a composite quantum system consisting of \( n \) qubits is the tensor product of the individual state spaces of each component system, i.e., \( \mathcal{H} = \mathcal{H}_1 \otimes \mathcal{H}_2 \otimes \dots \otimes \mathcal{H}_n \).
In the case of open quantum systems, the state is represented by a density operator \( \rhb = \sum_{k} p_k |\psi_k\rangle \langle \psi_k| \), where the indices \( k \) label the possible pure states, with \( p_k \geq 0 \) and \( \sum_k p_k = 1 \). A density operator \( \rhb \) is positive semi-definite and satisfies \( \mathrm{tr}(\rhb) = 1 \).
The dynamics of open quantum systems are often modeled using master equations, and the most general form of a master equation is the Lindblad equation \cite{lind1,lind2}, which guarantees both complete positivity and trace preservation. This equation governs the evolution of the density operator \( \rhb \) as follows,
\begin{equation}
    \label{eq:linb}
\frac{d\rhb}{dt} = -\frac{\imath}{\hbar} [\Hs, \rhb] + \sum_{l \in \mathbb{I}} \gamma_l \mathcal{D}(\Ls_l) \rhb,
\end{equation}
where \( \imath \) is the imaginary unit, \( \hbar \) is the reduced Planck constant, \(\Hs \) is the effective Hamiltonian, and \( \Ls_l \) are Lindblad operators. Each Lindblad operator \( \Ls_l \) is associated with a relaxation rate \( \gamma_l > 0 \), and the index \( l \) belongs to a set \( \mathbb{I} \). The dissipator \( \mathcal{D}[\Ls_l] \) acting on \( \rhb \) is defined as
\[
\mathcal{D}[\Ls_l] \rhb = \Ls_l \rhb \Ls_l^\dagger - \frac{1}{2} \left( \Ls_l^\dagger \Ls_l \rhb + \rhb \Ls_l^\dagger \Ls_l \right).
\]

\subsection{Graph Theory Essentials}

An undirected graph \( \mG = (\mV, \mE) \) consists of a set \( \mV = \{1, 2, \dots, n\} \) of nodes and a set \( \mE \) of edges, where an edge \( \{j,k\} \in \mE \) represents a connection between nodes \( j \) and \( k \), with \( j, k \in \mV\). A \emph{path} between two nodes \( v_1 \) and \( v_k \) in \( \mG \) is a sequence of distinct nodes \( v_1, v_2, \dots, v_k \), such that for every \( m = 1, \dots, k-1 \), there is an edge between \( v_m \) and \( v_{m+1} \). A pair of distinct nodes \( j \) and \( k \) is considered \emph{reachable} from each other if there exists a path connecting them. In addition, each node is assumed to be reachable from itself. The graph \( \mG \) is said to be \emph{connected} if every pair of distinct nodes in \( \mV \) is reachable from each other.

Next, define the weight matrix \( [a_{jk}] \in \mathbb{R}^{n \times n} \), where \( a_{jk} > 0 \) if \( \{j,k\} \in \mE \), and \( a_{jk} = 0 \) if \( \{j,k\} \notin \mE \). Then the \emph{Laplacian} of \( \mG \), denoted by \( \Qs \), is defined as,
\[
[\Qs]_{jk} = 
\begin{cases} 
-a_{jk} & \text{if } j \neq k, \\
\sum_{j=1}^n a_{jk} & \text{if } j = k.
\end{cases}
\]
It is well known that \( \Qs \) is always positive semi-definite, with \( p \) zero eigenvalues, where \( p \) denotes the number of connected components of \( \mG \).

\subsection{Blended Dynamic Theory for Classical Networks}

Consider a classical network of \( n \) nodes, indexed by \( \mV = \{1, \dots, n\} \), connected over a graph \( \mathrm{G}=(\mV, \mE) \), with the network dynamics being described by
\begin{equation}
    \label{eq:cla}
    \frac{{d\mathbf{x}}_j}{dt} = K_c \sum_{\{j,k\}\in\mathbb{E}} a_{jk} (\mathbf{x}_k - \mathbf{x}_j) + f_j(\mathbf{x}_j, t), \quad j \in \mV,
\end{equation}
where \( \mathbf{x}_j \in \mathbb{R}^m \) is the local state at node \( j \)  and \( K_c > 0 \) indicates the diffusion strength. The function \( f_j: \mathbb{R}^m \times \mathbb{R}_+ \rightarrow \mathbb{R}^m \) represents a local function governing the dynamics at node \( j \).

In \eqref{eq:cla}, the diffusive coupling term \( K_c \sum_{\{j,k\}\in\mathbb{E}} a_{jk} (\mathbf{x}_k - \mathbf{x}_j) \) drives the states at each node toward consensus, i.e., \( \mathbf{x}_j = \mathbf{x}_k \) for all \( j, k \in \mV \). If the gain \( K_c \) is sufficiently large, consensus is reached for any desired error at any desired time. Assuming that the consensus is reached and taking the average dynamic of each node, we then obtain the following corresponding 
\emph{blended dynamic system} for \eqref{eq:cla},
\begin{equation}
    \label{eq:cla_ble}
    \frac{d\sbf}{dt}= \frac{1}{n} \sum_{j \in \mV} f_j(\mathbf{s}, t),
\end{equation}
where $\sbf(t)\in \mathbb{R}^{m}$ and the initial condition is \( \mathbf{s}(0) = \frac{1}{n} \sum_{j \in \mV} \mathbf{x}_j(0) \). Consistent with this observation, several key results have been established, demonstrating the consistency between the blended dynamic system \eqref{eq:cla_ble} and the original system \eqref{eq:cla}. For instance, it has been shown that if the gain \( K_c \) is sufficiently large, the state \( \mathbf{x}_j(t) \) generated in \eqref{eq:cla} and the state \( \mathbf{s}(t) \) generated in \eqref{eq:cla_ble} will be approximately equal, with the error being arbitrarily small. 

According to the singular perturbation theory \cite{Khalil(2002)}, the difference between \eqref{eq:cla} and \eqref{eq:cla_ble} can be viewed as a singular perturbation when the diffusive coupling gain is sufficiently large. As a result, this theory is applicable to a variety of networks with diffusive coupling, as demonstrated in classical examples in \cite{Lee2022}, and will be extended to open quantum networks in this paper.

\subsection{Notations}
 The notation \( |\cdot\rangle \) denotes the Dirac notation, and the complex conjugate transpose of \( |y\rangle \) is represented by \( \langle y| \). The symbols \( \mathsf{1}_n \), \( \mathsf{0}_n \), and \( \Is_n \) refer to the one vector, zero vector, and identity operator in \( \mathbb{R}^n \), respectively. The Pauli operators \( \sigma_{x,y,z} \) and the raising/lowering operators \( \sigma_{\pm} \) are given as follows,
\[
\ba
&\sigma_x = \begin{bmatrix}
0 & 1 \\
1 & 0
\end{bmatrix}, \quad
\sigma_y = \begin{bmatrix}
0 & -\imath \\
\imath & 0
\end{bmatrix}, \quad
\sigma_z = \begin{bmatrix}
1 & 0 \\
0 & -1
\end{bmatrix}, \\
&\sigma_+ = \begin{bmatrix}
0 & 0 \\
1 & 0
\end{bmatrix}, \quad
\sigma_- = \begin{bmatrix}
0 & 1 \\
0 & 0
\end{bmatrix}.
\ea
\]
We use the notation \( \sigma^{(j)} \) to indicate that the operator \( \sigma \in \{ \sigma_x, \sigma_y, \sigma_z, \sigma_+, \sigma_- \} \) acts solely on the \( j \)-th qubit. For example, \( \sigma_+^{(2)} := \Is \otimes \sigma_+ \otimes \Is \otimes \dots \otimes \Is \).
 
 In addition, \( \mathrm{diag}(x_1, \dots, x_n) \) denotes a diagonal operator with \( x_j \) as the \( j \)-th diagonal element. For an operator \( \As \), its vectorization is represented by \( \mathrm{vec}(\As) \), and its inverse function is denoted by \( \mathrm{vec}^{-1} \). The symbols \( \oplus \), \( \|\cdot\| \), \( \|\cdot\|_{\rm F} \), and \( [\cdot,\cdot] \) represent the direct sum, the Euclidean norm, the Frobenius norm, and the commutator, respectively.  In addition, for a subspace $\mathbb{S} \subset \mathbb{C}^{n\times n}$, $\|\rhb\|_{\mathbb{S}}$ denotes the distance between the $\rhb \in \mathbb{C}^{n\times n}$ and $\mathbb{S}$,
i.e., $\|\rhb\|_{\mathbb{S}} := \inf_{\rhb'\in \mathbb{C}^{n\times n}} \|\rhb - \rhb'\|_{\rm F}$. 

\section{Diffusive and Open Quantum Networks}
\label{sec.dif}
We consider a quantum network consisting of $n$ qubits. The state of the network is described by a density operator $\rhb$, which is a Hermitian, positive semi-definite, and trace-one complex operator of dimension $2^n\times 2^n$ capturing the statistical mixture of different pure states.  There may be three types of interaction between the qubits:
\begin{itemize}
\item[(i)] The network Hamiltonian $\Hs$, which is a Hermitian operator describing the network energy and encoding the internal interactions. 
\item[(ii)] The environmental dissipations  described by the Lindblad operators $\mathcal{D}(\Ls_l)$ for $l\in\mathbb{I}$.  These environmental dissipations arise from the loss of energy, coherence, or information from the system to its surroundings (the “environment” or “bath”) due to unavoidable interactions between them. 
\item[(iii)] The diffusive couplings over an undirected and connected quantum graph \( \mG = (\mV, \mE) \) with weight matrix $[a_{jk}]$, where a swapping operator $\Us_{jk}$ along each $\{j,k\}\in \mE$ leads to the Lindblad interactions $ \mathcal{D}(\Us_{jk})$.  Here, the operator \( \Us_{jk} \)  exchanges the states of qubits \( j \) and \( k \) by
\[
\ba
&\Us_{jk} \big( \rhb_1 \otimes \dots \otimes \rhb_j \otimes \dots \otimes \rhb_k \otimes \dots \otimes \rhb_n \big) \Us_{jk}^\dagger \\=& \rhb_1 \otimes \dots \otimes \rhb_k \otimes \dots \otimes \rhb_j \otimes \dots \otimes \rhb_n ,
\ea
\]
for all \( \rhb_l \in \mathbb{C}^{2 \times 2} \) (\( l = 1, \dots, n \)) \cite{swap_operator}. The role of these swapping operators acts as a symmetrization force on $\rhb$, driving the reduced states of the qubits to a consensus state. Therefore, it serves as a generalization of the diffusive couplings in classical networks to the quantum domain. 
\end{itemize}


The time evolution of such a quantum network will be governed by the Lindblad master equation \eqref{eq:linb}, taking the following form
\begin{equation}
    \label{eq:obj}
    \frac{d\rhb}{dt} = -\frac{\imath}{\hbar} [\Hs, \rhb] + \sum_{l \in \mathbb{I}} \gamma_l \mathcal{D}(\Ls_l) \rhb + K_c \sum_{\{j,k\} \in \mE}   a_{jk}  \mathcal{D}(\Us_{jk}) \rhb,
\end{equation}
where \( K_c > 0 \) represents the gain of diffusive coupling.

Clearly, in its present form, (\ref{eq:obj}) defines a linear time-invariant system over the space of $\rhb$. While realistic open quantum networks can exhibit time-varying (\cite{openquantum}) and nonlinear (\cite{PhysRevResearch.6.013262}) dynamics, we focus on this simplified linear form to provide a clear and tractable analysis of the fundamental mechanisms. Because blended dynamics theory applies to general time-varying and nonlinear agent dynamics in the form of (\ref{eq:cla}), e.g., \cite{Lee2022}, all results presented here capture the essential behaviors and can, conceptually or analytically, be extended to nonlinear, higher-dimensional (non-qubit), and non-autonomous quantum networks.




\section{Emergence of Classical Behaviors}
\label{sec.cla}
In this section, we examine the dynamics of each reduced state in the quantum diffusive coupling master equation \eqref{eq:obj} and explore the classical collective behavior of the open quantum network. This is achieved by introducing blended reduced-state dynamics.
To address this issue, we consider the case where the Hamiltonian $\Hs$ and the operators $\Ls_l$ in \eqref{eq:obj} are separable, as outlined in the following assumption.

\begin{assumption}
\label{ass:sep}
    $\Hs$ and $\Ls_l$ are separable, i.e., $\Hs=\Hs_1\oplus \Hs_2\oplus\dots\oplus \Hs_n$, $\Ls_l=\Ls_{l,1}\oplus \Ls_{l,2}\oplus\dots\oplus \Ls_{l,n}$ for all $l\in \mathbb{I}$. In addition, the initial state of the system \eqref{eq:obj} is product state, i.e., $\rhb(0)=\otimes_{j\in\mV}\rhb_j(0)$ for some density operators $\rhb_j(0)$.
\end{assumption}

\subsection{Blended Reduced-state  Dynamics}
Under Assumption \ref{ass:sep}, the dynamics of the individual qubit of the quantum diffusive coupling master equation \eqref{eq:obj} can be expressed as
\begin{equation}
    \label{eq:partra}
    \frac{d\rhb_j}{dt} = -\frac{\imath}{\hbar}[\Hs_j, \rhb_j] + \sum_{l \in \mathbb{I}} \gamma_l\mathcal{D}(\Ls_{l,j}) \rhb_j+ K_c \sum_{\{j,k\}\in\mathbb{E}} a_{jk}  (\rhb_k - \rhb_j),
\end{equation}
for all $j\in\mV$, where $\rhb_j := \mathrm{Tr}_{\otimes_{k\neq j} \mathcal{H}_k}(\rhb)$ is the \emph{reduced state} of qubit $j$, defined by the partial trace over the space of the remaining qubits.
In \eqref{eq:partra}, the term \( K_c \sum_{\{j,k\} \in \mathbb{E}} a_{jk} (\rhb_k - \rhb_j) \) arises from the diffusive coupling term in \eqref{eq:obj}, which drives the reduced states toward consensus, i.e., \( \rhb_j = \rhb_k \) for all \( j, k \in \mathbb{V} \), with arbitrarily small error as the coupling gain \( K_c \) becomes sufficiently large. The blended dynamics of the reduced qubit states can then be established as a direct extension of their classical counterpart.
\begin{definition} [{Blended Reduced-state Dynamics}]
    \label{def:blen1}
    The blended reduced-state dynamics of \eqref{eq:obj} is defined as 
\begin{equation}
    \label{eq:blen1}
   \frac{d \rhb_{\rm b}}{dt} = -\frac{\imath}{\hbar} [\bar{\Hs}, \rhb_{\rm b}] + \sum_{l \in \mathbb{I}} \gamma_l\mathcal{D}(\bar{\Ls}_l) \rhb_{\rm b},
\end{equation}
with 
the initial condition $\rhb_{\rm b}(0) = \frac{1}{n} \sum_{j \in \mV} \rhb_j(0)$, where $\bar{\Hs} = \frac{1}{n} \sum_{j \in \mV} \Hs_j$ and $\bar{\Ls}_l = \frac{1}{n} \sum_{j \in \mV} \Ls_{l,j}$. Furthermore, we denote the evolution described by \eqref{eq:blen1} as $\bar{\Lc} \rhb_{\rm b} := \dfrac{d \rhb_{\rm b}}{dt}$.
\end{definition}

\subsection{Clustering toward a Common Equilibrium}
 
Considering a special and important case where \eqref{eq:blen1} is \emph{relaxing}, meaning that there exists a steady state $\rhb_{\rm r}$ such that
\[
    \lim_{t \to \infty} \rhb_{\rm b}(t) = \rhb_{\rm r}, \quad \forall\rhb_{\rm b}(0),
\]
we establish the following theorem.

\begin{theorem}
\label{cor:rex}
    Let Assumptions \ref{ass:sep}  hold. Assume that the blended reduced-state dynamics \eqref{eq:blen1} is relaxing.  Then for any given $\eta>0$, there exists some $K_c^\ast>0$
    such that for any $K_c\geq K_c^\ast$, it holds along \eqref{eq:obj} that
   \begin{equation}
       \label{eq:thm1}
\ba
&\max_{j\in\mV}\|\rhb_{j}(t)-\rhb_{\rm r}\|_{\rm F}\leq Ce^{-\frac{\mu t}{2}},\ &\quad&\forall 0\leq t\leq \frac{2}{\mu} \ln \left( \frac{C}{\eta} \right),\\&\max_{j\in\mV}\|\rhb_{j}(t)-\rhb_{\rm r}\|_{\rm F}\leq \eta,  &&\forall t\geq \frac{2}{\mu} \ln \left( \frac{C}{\eta} \right).
\ea
   \end{equation}
    where \( C =(1+\sqrt{2})\left( \sqrt{\sum_{j \in \mathbb{V}} \|\rhb_j(0) - \bar{\rhb}(0)\|_{\rm F}^2} + L_c\right) \) for some $L_c>0$, $\bar\rhb(0)=\frac{1}{n}\sum_{j \in \mathbb{V}} \rhb_j(0)$, and $\mu$ is the absolute of the minimal nonzero real part of eigenvalues of $\bar{\Lc}$. 

\end{theorem}
The proof of Theorem \ref{cor:rex} can be seen in Appendix \ref{app:thm1}. {The lower bound for coupling gain $K_c^\ast$ in Theorem \ref{cor:rex} is explicitly given by
    \[K_c^\ast=\max\left\{ \frac{\mu}{\lambda_{\rm m}(\Qs)}, \frac{2(1+\sqrt{2})D}{\eta} \right\},\] 
    where
\(
    D =  \frac{L_c \|\Lc\|}{\lambda_{\rm m}(\Qs)} \left( \frac{1}{n} \sqrt{\sum_{j \in \mathbb{V}} \|\rhb_j(0) - \bar{\rhb}(0)\|_{\rm F}^2} + \frac{\|\Lc\|}{\mu} \right)+\frac{n \|\Lc\|}{\lambda_{\rm m}(\Qs)} ,
\) $\Lc=\bigoplus_{j \in \mathbb{V}} \mathcal{L}_j$ with \(\Lc_j\) being defined by \( \Lc_j \rhb_j := -\frac{\imath}{\hbar}[\Hs_j, \rhb_j] + \sum_{l \in \mathbb{I}}\gamma_l \mathcal{D}(\Ls_{l,j}) \rhb_j \) for any \(\rhb_j \in \mathbb{C}^{2 \times 2}\), and \(\lambda_{\rm m}(\Qs)\) is the minimal nonzero eigenvalue of \(\Qs\), the Laplacian of the graph $\mG$.} Theorem \ref{cor:rex} demonstrates that, with sufficiently large 
$K_c$, the reduced states in \eqref{eq:obj} will converge to an arbitrarily small $\eta$-neighborhood, of the steady state of \eqref{eq:blen1}, with exponential convergence rates.

\medskip

\begin{figure*}[t]
   \centering
    \subfigure[$\eta=0.1,K=6$]{\includegraphics[width=5.8cm]{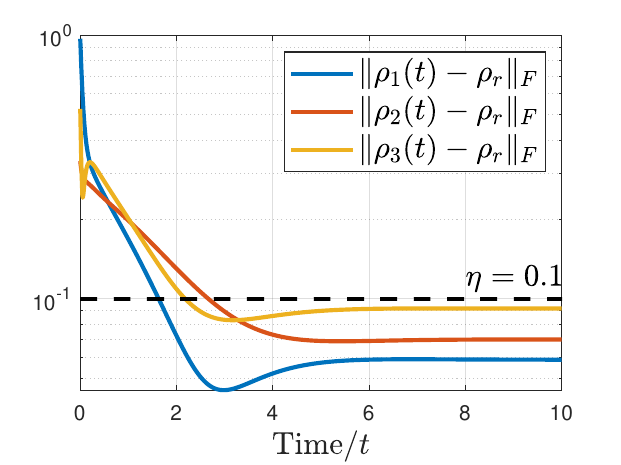}}
    \subfigure[$\eta=0.05,K=15$]{\includegraphics[width=5.8cm]{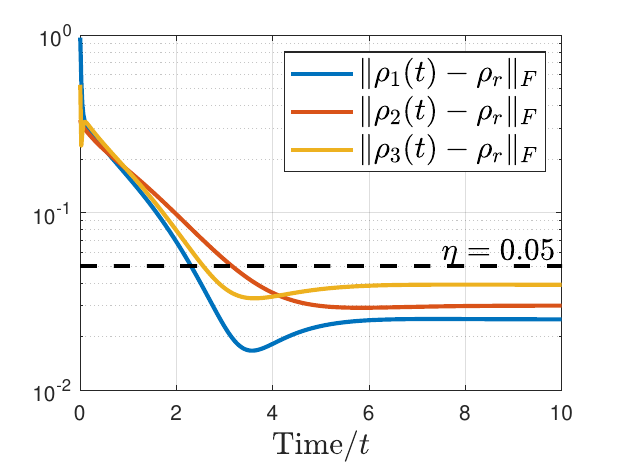}}
    \subfigure[$\eta=0.01,K=70$]{\includegraphics[width=5.8cm]{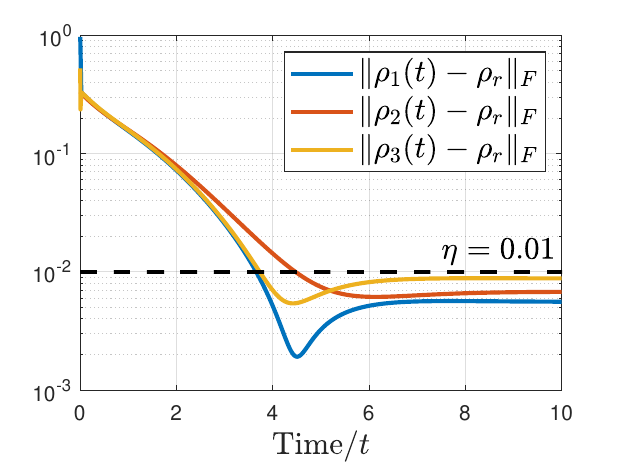}}
    \caption{The time evolution of the Frobenius distance between the reduced states $\rhb_1(t)$, $\rhb_2(t)$, and $\rhb_3(t)$ in \eqref{eq:sim1} and the steady state $\rhb_{\rm r}$ in \eqref{eq:blen_sim1} with varying diffusive coupling gain $K_c$ for given $\eta$.
}
    \label{fig.sim1}
\end{figure*}

\noindent{\bf Example 1}.  We consider 3 qubits connected on a complete quantum graph, which describes a dissipative spin system studied in~\cite{sim1}. In this system, the qubits $1$, $2$, and $3$ are subject to different local dynamics. Qubit~$1$ is driven by a coherent Hamiltonian $\sigma_z$, qubit~$2$ undergoes lowering damping with Lindblad operator $\sigma_{-}$, and qubit~$3$ undergoes raising damping with Lindblad operator $\sqrt{3}\sigma_{+}$. In this setting, the master equation~\eqref{eq:obj} becomes
\begin{equation}
    \label{eq:sim1}
    \ba
    \frac{d\rhb}{dt} = &-\frac{\imath}{\hbar} [\sigma_z^{(1)}, \rhb] +  \mathcal{D}(\sigma_-^{(2)}) \rhb  + \mathcal{D}(\sqrt{3}\sigma_+^{(3)}) \rhb + K_c \sum_{\{j,k\} \in \mE}  \mathcal{D}(\Us_{jk}) \rhb,
    \ea
\end{equation}
where the initial state is set as $\rhb(0) = |0\rangle\langle 0| \otimes \frac{1}{2}I \otimes |1\rangle\langle 1|$.
Accordingly, the associated blended reduced-state dynamics is given by
\begin{equation}
    \label{eq:blen_sim1}
    \frac{d\rhb_{\rm b}}{dt} = -\frac{\imath}{\hbar} \left[\frac{1}{3} \sigma_z, \rhb_{\rm b} \right] + \frac{1}{3} \left( \mathcal{D}(\sigma_-) \rhb_{\rm b} + \mathcal{D}(\sqrt{3}\sigma_+) \rhb_{\rm b} \right),
\end{equation}
with the initial condition $\rhb_{\rm b}(0) = \frac{1}{2}\Is$.
It is straightforward to verify that Assumption~\ref{ass:sep} holds, and that~\eqref{eq:blen_sim1} is relaxing with a unique steady state
\begin{equation*}
\rhb_{\rm r} = \begin{bmatrix}
\frac{2}{3} & -\frac{i}{6} \\
\frac{i}{6}& \frac{1}{3}
\end{bmatrix}.
\end{equation*}

In Figure~\ref{fig.sim1}, we present the evolution of the Frobenius distance between each reduced state $\rhb_j(t)$ and the steady state $\rhb_{\rm r}$, denoted by $\|\rhb_1(t)-\rhb_{\rm r}\|_{\rm F}$, $\|\rhb_2(t)-\rhb_{\rm r}\|_{\rm F}$, and $\|\rhb_3(t)-\rhb_{\rm r}\|_{\rm F}$, for various values of the diffusive coupling gain $K_c$.
It can be observed that the reduced states converge exponentially to an $\eta$-neighborhood of $\rhb_{\rm r}$ for any given $\eta > 0$, provided that $K_c$ is sufficiently large. 
\hfill$\square$

   Theorem \ref{cor:rex} discusses the case where \eqref{eq:blen1} is relaxing. For more general case where \eqref{eq:blen1} has an invariant subspace $\mathbb{S}\subset\mathbb{C}^{2\times2}$, we can conclude that Theorem \ref{cor:rex} holds with \eqref{eq:thm1} becoming
    \begin{equation}
    \label{eq:remark}
    \ba
     E(t) &\leq C e^{-\frac{\mu t}{2}}, \quad  &&\forall 0 \leq t \leq \frac{2}{\mu} \ln \left( \frac{C}{\eta} \right), \\
    E(t) &\leq \eta,  &&\forall t \geq \frac{2}{\mu} \ln \left( \frac{C}{\eta} \right),
    \ea
    \end{equation}
    where $E(t) := \max_{j\in\mV} \left\{\left\|\rhb_j(t)-\bar\rhb(t) \right\|_{\rm F} + \|\bar\rhb(t)\|_\mathbb{S}\right\}$ and $\bar\rhb(t):=\frac{1}{n}\sum_{j\in\mV}\rhb_j(t)$. This indicates that for sufficiently large $K_c$, the reduced states generated in \eqref{eq:obj} will converge to the invariant subspace of the blended dynamic system \eqref{eq:blen1} and will reach the neighbourhood of consensus with an arbitrarily small error.

\subsection{Converging to a Common Trajectory}

The following theorem establishes the consistency between the evolution of the reduced states $\rhb_j(t)$ and that of $\rhb_{\rm b}(t)$ in \eqref{eq:blen1}. 

\begin{theorem}
\label{thm:blen1_o}
Let Assumption \ref{ass:sep}  hold and $\rhb_{\rm b}(t)$ be generated by \eqref{eq:blen1}.  Then for any given $\eta>0$ and $T_2\geq T_1>0$, there exists some $K_c^\ast>0$ such that for any $K_c\geq K_c^\ast$, it holds along \eqref{eq:obj} that
\[
\ba
\max_j\|\rhb_j(t)-\rhb_{\rm b}(t)\|_{\rm F}\leq\eta,\quad \forall t\in[T_1,T_2].
\ea
\]
\end{theorem}

The proof of Theorem \ref{thm:blen1_o} can be seen in Appendix \ref{app:thm2}. The value of $K_c^\ast$  in Theorem \ref{thm:blen1_o} is given by
\[
 K_{c}^\ast = \max\left\{\frac{1}{T_1\lambda_{\rm m}(\Qs)}\ln\left(\frac{2C}{\eta}\right), \frac{2D(T_2)}{\eta}\right\},
\]
where
$C=\sqrt{\sum_{j\in\mV}\|\rhb_j(0)-\bar{\rhb}(0)\|_{\rm F}^2}$, $D(T_2):=\frac{n\|\Lc\|}{\lambda_{\rm m}(\Qs)}+\frac{f(T_2)\|\Lc\|}{n\lambda_{\rm m}(\Qs)}\sqrt{\sum_{j\in\mV}\|\rhb_j(0)-\bar{\rhb}(0)\|_{\rm F}^2}$, and \( f(T_2) = M \sum_{j=0}^{m} T_2^j \) for some \( m,M > 0 \).

Notably, compared with Theorem \ref{cor:rex}, in Theorem \ref{thm:blen1_o}, the relaxing condition is not required. Theorem \ref{thm:blen1_o} demonstrates that the evolution of 
$\rhb_j(t)$ over time tends to be close to the blended reduced-state dynamics \eqref{eq:blen1} over a finite time horizon with strong diffusive couplings.  That is, \( \rhb_j(t) \) will still converge to the \( \eta \)-neighborhood of \( \rhb_{\rm b}(t) \), even if the latter exhibits dynamic orbital solutions instead of stabilizing to an equilibrium. The results of Theorems 1 and 2 are consistent with those in classical blended dynamics theory, specifically with Theorem 2 in \cite{Lee2022}.


\medskip

\noindent{\bf Example 2}.   For the same quantum graph discussed in Example 1, we consider a system of three qubits driven by the Hamiltonian \( \Hs = \sigma_x^{(1)} + \sigma_y^{(2)} + \sigma_z^{(3)} \) and without dissipation, i.e., with Lindblad operators \( \Qs = 0 \)~\cite{sim2}. The initial state is set to \( \rhb(0) = |0\rangle\langle 0| \otimes \frac{1}{2}(\Is+|0\rangle\langle 1|+|1\rangle\langle 0|) \otimes |1\rangle\langle 1| \). The original system and the corresponding blended reduced-state dynamics are described as follows,
\begin{equation}
    \label{eq:sim2}
    \frac{d\rhb}{dt} = -\frac{\imath}{\hbar} \left[ \sigma_x^{(1)} + \sigma_y^{(2)} + \sigma_z^{(3)},\, \rhb \right],
\end{equation}
\begin{equation}
    \label{eq:blen_sim2}
    \frac{d\rhb_{\rm b}}{dt} = -\frac{\imath}{\hbar} \left[ \frac{1}{3}(\sigma_x + \sigma_y + \sigma_z),\, \rhb_{\rm b} \right],
\end{equation}
with the initial condition for~\eqref{eq:blen_sim2} given by \( \rhb_{\rm b}(0) = \frac{1}{2}\Is+\frac{1}{6}|0\rangle\langle 1|+\frac{1}{6}|1\rangle\langle 0|\).

\begin{figure*}[t]
   \centering
    \subfigure[$T_2=4,K=60$]{\includegraphics[width=5.8cm]{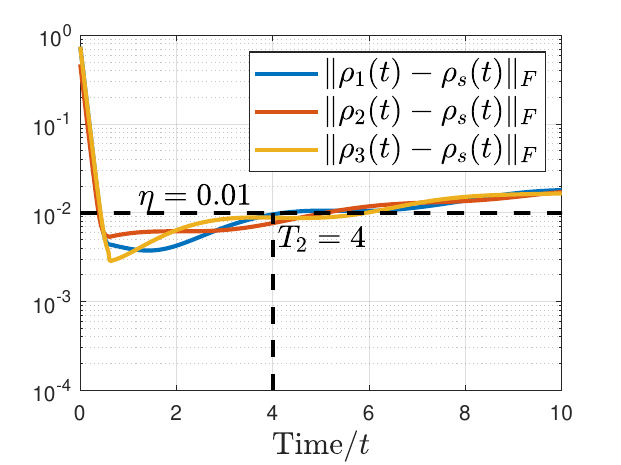}}
    \subfigure[$T_2=6,K=80$]{\includegraphics[width=5.8cm]{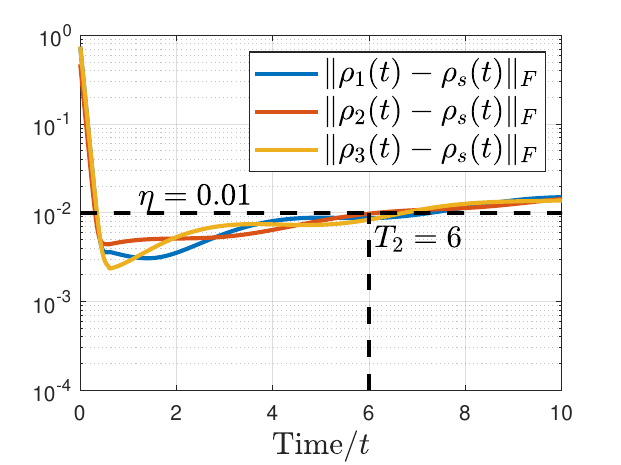}}
    \subfigure[$T_2=8,K=100$]{\includegraphics[width=5.8cm]{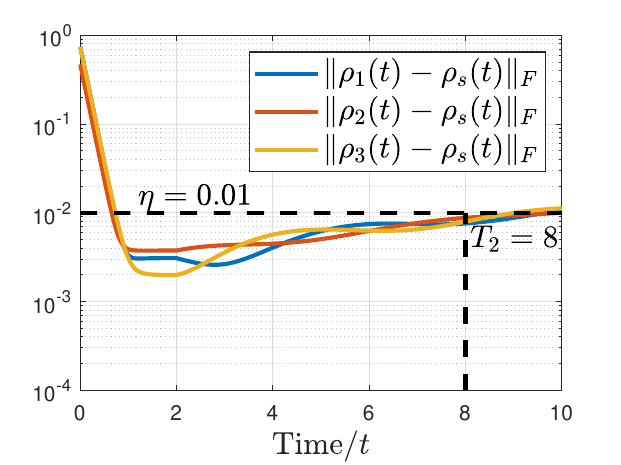}}
    \caption{The time evolution of Frobenius distance between the reduced states $\rhb_1(t)$, $\rhb_2(t)$, and $\rhb_3(t)$ in \eqref{eq:sim2} and the density operator $\rhb_{\rm b}(t)$ in \eqref{eq:blen_sim2}  with varying diffusive coupling gain $K_c$ for given $T_2$ and $\eta=0.01$.}
    \label{fig.sim2}
\end{figure*}

\begin{figure}[t]
    \centering
    \subfigure{\includegraphics[width=8cm]{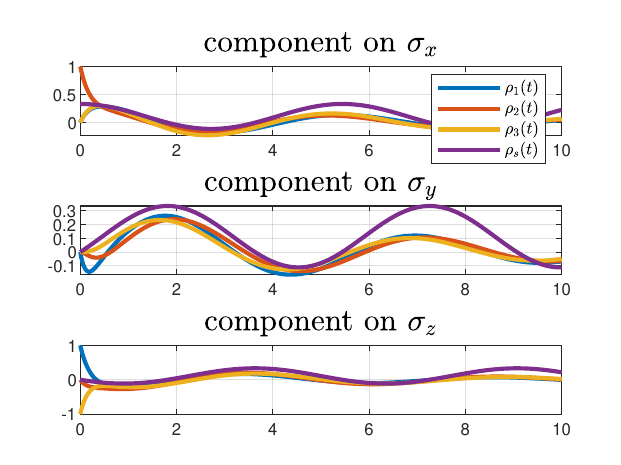}}\quad
    \subfigure{\includegraphics[width=8cm]{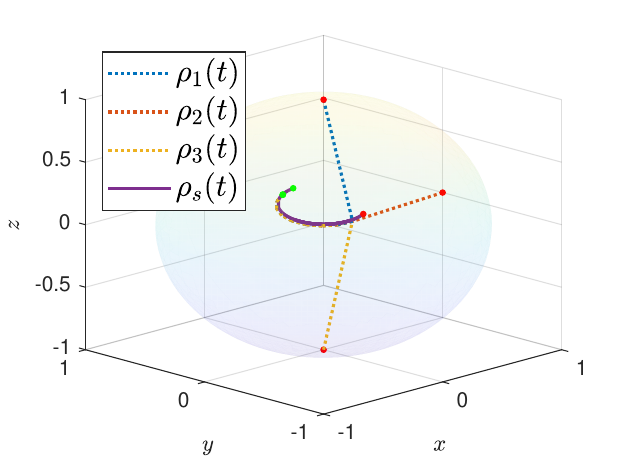}}
    \caption{The evolutions of components of Bloch vectors of  $\rhb_1(t),\rhb_2(t),\rhb_3(t)$ in \eqref{eq:sim2} and $\rhb_{\rm b}(t)$ in  \eqref{eq:blen_sim2} with $K_c=40$. The red dots and the green dots denote the initial states and final states of the trajectories, respectively.}
    \label{fig.sim2_2}
\end{figure}

It can be concluded that both the reduced states of~\eqref{eq:sim2} and the density operator \( \rhb_{\rm b}(t) \) in~\eqref{eq:blen_sim2} exhibit orbital behavior. The time evolution of the Frobenius distances between each reduced state and \( \rhb_{\rm b}(t) \), i.e., \( \|\rhb_1(t) - \rhb_{\rm b}(t)\|_{\rm F} \), \( \|\rhb_2(t) - \rhb_{\rm b}(t)\|_{\rm F} \), and \( \|\rhb_3(t) - \rhb_{\rm b}(t)\|_{\rm F} \), is shown in Fig.~\ref{fig.sim2}.
For a more intuitive comparison, Fig.~\ref{fig.sim2_2} shows the evolution of the Bloch vectors of the reduced states \( \rhb_1(t) \), \( \rhb_2(t) \), \( \rhb_3(t) \), and the blended state \( \rhb_{\rm b}(t) \) for \( K_c = 40 \)\footnote{Each single-qubit density operator can be uniquely represented as \( \rhb_j(t) = \frac{1}{2}I + x(t)\sigma_x + y(t)\sigma_y + z(t)\sigma_z \), where \( \rb(t) = [x(t), y(t), z(t)] \) is the Bloch vector.}.
From both figures, it is observed that each reduced state \( \rhb_j(t) \) remains within an \( \eta \)-neighborhood of \( \rhb_{\rm b}(t) \) after a finite time \( T_1 \), although the error increases over time. For any fixed \( \eta \) and finite duration \( T_2 \), the reduced states remain within the \( \eta \)-neighborhood of \( \rhb_{\rm b}(t) \) up to time \( T_2 \), provided that \( K_c \) is sufficiently large.  \hfill $\square$

\section{Emergence of Coherent  Behaviors}
\label{sec.eme}
In the previous section, we explored the case where the evolutions of the reduced states can be obtained individually by assuming that \( \Hs \) and \( \Ls_l \) are separable. In this section, we analyze the general case where they are inseparable. To this end, we utilize the induced graph of the quantum graph \( \mG \) to propose the blended coherent dynamics, thus studying the coherent collective behaviors of the open quantum network.

\subsection{The Induced Graph}
We recall the concept of the induced graph introduced in \cite{Shi2016Reaching}.
Assuming that the density operator \( \rhb \) of the quantum network is defined with respect to the basis \( \{|l\rangle\} \) for \( l \in \mathbb{B} := \{0, 1, \dots, 2^n - 1\} \) of the \( n \)-dimensional Hilbert space \( \mathcal{H}^{\otimes n} \), we have
$$
\rhb = \sum_{l,l' \in \mathbb{B}} [\rhb]_{ll'} |l\rangle \langle l'|,
$$
where \( [\rhb]_{ll'} \) denotes the \( (l+1) \)-th row and \( (l'+1) \)-th column element of \( \rhb \). Next, we index the basis by \( \mV_{\mathrm{d}} = \{|l\rangle\langle l'|\mid l, l' \in \mathbb{B}\} \).  
Define the mapping \( \mathsf{S}_{jk}(l): \{0, 1\}^n \to \{0, 1\}^n \), which exchanges the \( j \)-th and \( k \)-th bits of the binary number \( l \in \mathbb{B} \), where \( j, k = 1, 2, \dots, n \). Then, we obtain the edge set
\[
\mE_{\mathrm{d}} = \left\{ \{|l\rangle \langle l'|, |\mathsf{S}_{jk}(l)\rangle \langle \mathsf{S}_{jk}(l')|\} \mid l, l' \in \mathbb{B}, \{j, k\} \in \mE \right\},
\]
where the weight of the edge \( \{|l\rangle \langle l'|, |\mathsf{S}_{jk}(l)\rangle \langle \mathsf{S}_{jk}(l')|\} \) is \( a_{jk} \) in the weight matrix of \( \mG \).  
For the quantum graph $\mG(\mV,\mE)$, its 
induced graph is now given by $\mG_{\mathrm{d}}(\mV_{\mathrm{d}},\mE_{\mathrm{d}})$.

\begin{figure}[t] \centering
 \includegraphics[width=8cm]{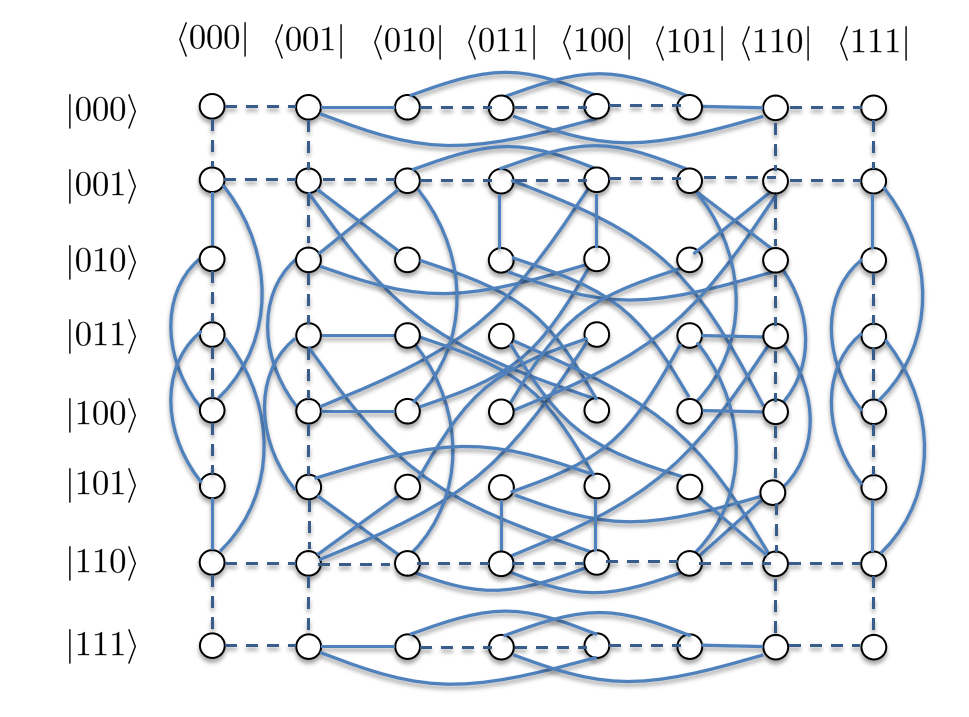}
    \caption{The illustration of the induced graph for a 3-qubit connected network governed by \eqref{eq:obj_in}. In (a), the dashed lines represent the coupling between the induced graph nodes, arising from the term \( \As_{\mathrm{d}} \tilde{\yb} \) (part), while the solid lines represent the diffusive coupling due to the term \( \Qs_{\mathrm{d}} \tilde{\yb} \). }
    \label{fig.reduced}
\end{figure}

  Fig.~\ref{fig.reduced} shows the induced graph of a quantum complete graph with three qubits. It can be observed that \( \mG_{\mathrm{d}} \) is undirected. In addition, we assume that there are \( p \) connected components in \( \gc \), which implies that the Laplacian of the induced graph, denoted by  \( \Qs_{\mathrm{d}} \), is semi-definite with \( p \) zero eigenvalues by \cite{mag}.
  
Next, we introduce the following notations to facilitate the subsequent analysis. Let \( \mG_{m}(\mV_{m}, \mE_m) \) denote the connected component of \( \mG_{\mathrm{d}} \), where \( m = 1, 2, \dots, p \). We know that \( \{\mathbf{p}_1, \mathbf{p}_2, \dots, \mathbf{p}_p\} \) forms a set of orthonormal basis vectors in the null space of \( \Qs_{\rm d} \), with each \( \mathbf{p}_m \in \mathbb{R}^{4^n} \) for \( m = 1, 2, \dots, p \), and is defined as follows,
\[
\ba
\ [\mathbf{p}_m]_k &= 0, \quad \text{for } k \notin \mV_m, \\
[\mathbf{p}_m]_k &= \frac{1}{|\mV_m|}, \quad \text{for } k \in \mV_m.
\ea
\]
For an operator \( \As \in \mathbb{C}^{4^n \times 4^n} \) and a vector \( \mathbf{x} \in \mathbb{C}^{4^n} \), their projections onto the null space of \( \Qs_{\mathrm{d}} \) are defined as
\[
\mathrm{Proj}^\perp_{\Qs_{\mathrm{d}}}(\As) := \Pb_{\rm d}^{\top} \As \Pb_{\rm d}, \quad \mathrm{Proj}^\perp_{\Qs_{\mathrm{d}}}(\mathbf{x}) := \Pb_{\rm d}^{\top} \mathbf{x},
\]
where \( \Pb_{\rm d} := [\mathbf{p}_1, \mathbf{p}_2, \dots, \mathbf{p}_p] \in \mathbb{R}^{4^n \times p} \). Furthermore, the inverse projection of \( \mathbf{x}_{\rm d} \in \mathbb{C}^{p} \) is given by
\[
\overline{\mathrm{Proj}}^\perp_{\Qs_{\mathrm{d}}}(\mathbf{x}_{\rm d}) := \Pb_{\rm d} \mathbf{x}_{\rm d}.
\]

Recall that a permutation $\pi$ of the set $\mathbb{V}$ is a bijective map
from $\mathbb{V}$ onto itself. The set of all permutations of $\mathbb{V}$  forms a group,
called the $n$'th permutation group and denoted by $\mathbb{P} = \{\pi\}$.
There are $n!$ elements in $\mathbb{P}$.
Given $\pi \in \mathbb{P}$, we define a unitary operator $\mathsf{U}_\pi$ over $\mathcal{H}^{\otimes n}$ by
\[
\mathsf{U}_\pi\big( |q_1\rangle \otimes \cdots \otimes |q_n\rangle \big) = |q_{\pi(1)}\rangle \otimes \cdots \otimes |q_{\pi(n)}\rangle,
\]
where $q_i \in \{0,1\}$ for all $i \in\mV$. 
Define an operator over the density operators of $\mathcal{H}^{\otimes n}$, $\mathcal{P}_\ast$,   as the {\emph{permutation-invariant projection} of $\rhb$ (see \cite{swap_operator})
\[
\mathcal{P}_\ast(\rhb) = \frac{1}{n!} \sum_{\pi \in P} \mathsf{U}_\pi \,\rhb\, \mathsf{U}_\pi^\dagger.
\]
Letting \( \tilde{\yb} := \mathrm{vec}(\rhb) \in \mathbb{C}^{4^n} \), we have 
\begin{equation}
    \label{eq:proandinv}
\mathsf{P}_{\mathrm{d}}\mathsf{P}_{\mathrm{d}}^\top\tilde{\yb} = \mathrm{vec} (\mathcal{P}_\ast(\rhb)).
\end{equation}

\subsection{Blended Coherent Dynamics }

The quantum network dynamics \eqref{eq:obj} can be rewritten as the following vectorized form
\begin{equation}
\label{eq:obj_in}
    \frac{d}{dt} \tilde{\yb} = \As_{\mathrm{d}} \tilde{\yb} + K_c \Qs_{\mathrm{d}} \tilde{\yb},
\end{equation}
where \( \As_{\mathrm{d}} = \frac{\imath}{\hbar}(\Hs \otimes \Is_{2^n} + \Is_{2^n} \otimes \Hs) + \sum_{l \in \mathbb{I}} \gamma_l
(\Ls_{l} \otimes \Ls_{l}^\dagger 
- \frac{1}{2} \Ls_{l}^\dagger \Ls_{l} \otimes \Is_{2^n} 
- \frac{1}{2} \Is_{2^n} \otimes \Ls_{l}^{\top} \Ls_{l}^\dagger) \).

The evolution of \( \tilde{\yb}(t) \) in \eqref{eq:obj_in} can be interpreted as the dynamics of the nodes in the induced graph \( \mG_{\mathrm{d}} \), where the state of each node corresponds to the respective component of \( \tilde{\yb}(t) \). In \eqref{eq:obj_in}, the first term \( \As_{\mathrm{d}} \tilde{\yb} \) represents the coupling between the nodes, derived from the effective Hamiltonian \( \Hs \) and the Lindblad operators \( \Ls_l \) in \eqref{eq:obj}, while the second term \( K_c \Qs_{\mathrm{d}} \tilde{\yb} \) captures the diffusive coupling as described in \eqref{eq:obj}. 

\begin{figure*}[t]
    \centering
     \includegraphics[height=6cm]{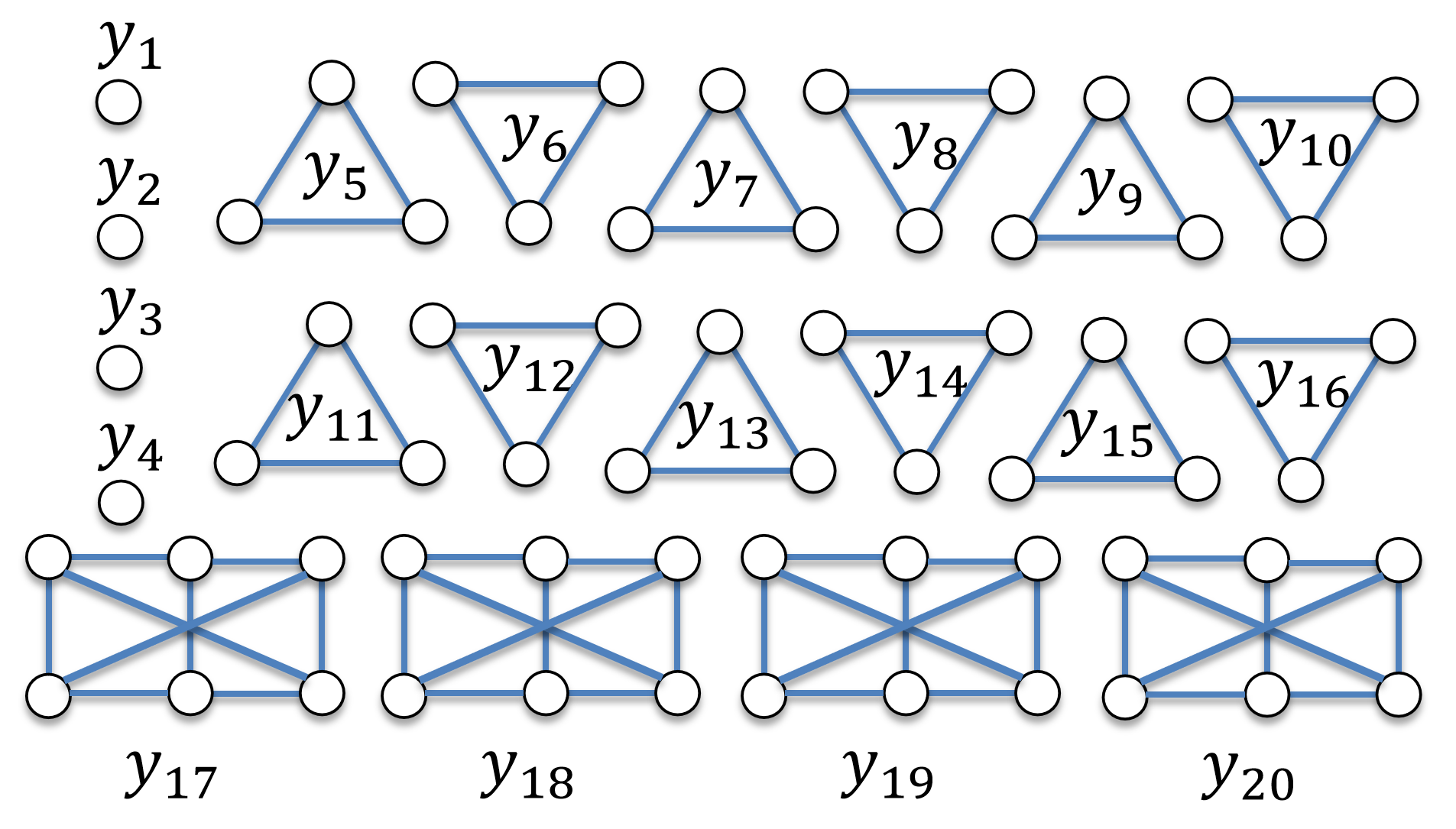}
    \caption{The illustration of the blended coherent dynamics:   once the nodes within each connected component have reached consensus, the coupling \( \As_{\mathrm{d}} \tilde{\yb} \) can be established between the connected components.}
    \label{fig.illu2}
\end{figure*}

Moreover, it can be seen that if \( K_c \) in \eqref{eq:obj} is sufficiently large, \( \tilde{\yb}(t) \) will approach the null space of \( \Qs_{\mathrm{d}} \), where the states within each connected component of \( \mG_{\mathrm{d}} \) reach consensus. Thus, following the same reasoning used to derive \eqref{eq:blen1}, we may {\em assume} that the states of the nodes within each connected component \( \mG_{m} \), for \( m = 1, 2, \dots, p \), have {\em reached consensus}, denoted by \( y_{m} \in \mathbb{C} \), as illustrated in Fig.~\ref{fig.illu2}. In this case, we have,
\begin{equation}
\label{eq:yyd}
\tilde{\yb}=\overline{\mathrm{Proj}}^\perp_{\Qs_\mathrm{d}}(\yb_{\rm d}),\end{equation}
where  \( \yb_{\rm d} := [|\mV_1|y_1; |\mV_2|y_2; \dots; |\mV_p|y_p] \in \mathbb{R}^{p} \). In addition, noting that \( \pb_{\rm d}^\top \pb_{\rm d} = \Is \), we have
\begin{equation}
\label{eq:ydy}
    \yb_{\rm d}={\mathrm{Proj}}^{\perp}_{\Qs_\mathrm{d}}(\tilde{\yb}).
    \end{equation} 

From \eqref{eq:yyd} and \eqref{eq:ydy}, it follows that when the states of the nodes within each connected component have reached consensus, the coupling of \( \tilde{\yb} \) in \eqref{eq:obj_in} can be reformulated in terms of the states \( \yb_{\rm d} \). Specifically, substituting \eqref{eq:yyd} and \eqref{eq:ydy} into \eqref{eq:obj_in}, we obtain a direct expression for the blended coherent dynamics as follows.

\begin{definition}
    [Blended Coherent Dynamics]
    \label{def:blen_in}
    The blended coherent dynamics of the quantum network \eqref{eq:obj} is defined as
\begin{equation}
\label{eq:blen_in}
    \frac{d}{dt} \yb_{\mathrm{d}} = \mathrm{Proj}^\perp_{\Qs_\mathrm{d}}(\As_{\mathrm{d}})\yb_{\mathrm{d}},
\end{equation}
with initial condition \( \yb_{\mathrm{d}}(0) =\mathrm{Proj}^\perp_{\Qs_\mathrm{d}}(\mathrm{vec}(\rhb(0)))\in\mathbb{C}^p \).
\end{definition}


\subsection{Coherent Orbit Attraction}

We define $\tilde{\yb}_{\mathrm{d}}:=\overline{\mathrm{Proj}}^\perp_{\Qs_\mathrm{d}}(\yb_{\rm d})\in\mathbb{C}^{4^n}$ as the inverse projection of $\yb_{\mathrm{d}}$. Then we have the following theorem describing the connection between \eqref{eq:obj} and \eqref{eq:blen_in}.

\begin{theorem}
\label{thm:blen1_in2}
Let $\yb_{\mathrm{d}}(t)$ be generated by \eqref{eq:blen_in}. Then for any $\eta>0,T_2\geq T_1>0$, there exists some $K_c^\ast>0$ such that for any $K_c\geq K_c^\ast$, $\rhb(t)$ generated in \eqref{eq:obj} satisfies
\[
\ba
{
\|\rhb(t)-\mathrm{vec^{-1}}(\tilde{\yb}_{\mathrm{d}}(t))\|_{\rm F}\leq\eta,\quad \forall t\in[T_1,T_2]}.
\ea
\]
\end{theorem}

The proof of Theorem \ref{thm:blen1_in2} can be seen in Appendix \ref{app:thm3}. Specifically, the value of $K_c^\ast$ in Theorem \ref{thm:blen1_in2} is given by
\[
K_{c}^\ast =\max\left\{ \frac{1}{T_1\lambda_{\rm m}(\Qs_{\rm d})}\ln\left(\frac{2C}{\eta}\right), \frac{2D(T_2)}{\eta}\right\},
\]
where $C=\|{\rhb}(0)-\mathcal{P}_\ast(\rhb(0))\|_{\rm F}$, $D(T_2)= \frac{\|\As_{\rm d}\|}{\lambda_{\rm m}(\Qs_{\rm d})}+\frac{f_{\mathrm{d}}(T_2) \|\As_{\mathrm{d}}\|}{\lambda_{\mathrm{m}}(\Qs_{\mathrm{d}})}\left( C   + T_2{\|\As_{\rm d}\|}\right)$, \(\lambda_{\rm m}(\Qs_{\rm d})\) is the minimal nonzero eigenvalue of \(\Qs_{\rm d}\), and \(f_{\mathrm{d}}(T_2) = M_{\mathrm{d}} \sum_{j=0}^{m_{\rm d}} T_2^j\) for some \(m_{\rm d},M_{\mathrm{d}}  >0\). 
Theorem \ref{thm:blen1_in2} suggests that for sufficiently large \( K_c \),  \( \rhb(t) \) will be consistent with the induced blended dynamic system in \eqref{eq:blen_in}, with any arbitrarily small error. In addition, Theorem \ref{thm:blen1_in2} is consistent with the conclusion in the classical blended dynamics theory for the case of rank-deficient coupling, specifically with Theorem 2 in \cite{LEE2020108952}.

In Theorems \ref{thm:blen1_o} and \ref{thm:blen1_in2}, the blended dynamic system and the original one can only be consistent  within a {\em finite} time \( T_2 \). This is because, in the invariant subspace of the main equation, \( \rho(t) \) may have an orbital solution, and a finite perturbation could cause the solution to exhibit {\em polynomially growing} error over time (see \( f(t) \) in Appendices \ref{app:thm2} and \ref{app:thm3}). On the other hand, in particular, when there is no orbital solution in the invariant subspace (e.g., when there is only a stable solution \( \rho_r \) in the subspace), the consistency between the two systems is not constrained by any time upper bound, consistent with Theorem \ref{cor:rex}.


In Theorem \ref{thm:blen1_in2}, we demonstrate the consistency between \eqref{eq:blen_in} and \eqref{eq:obj} within any finite time. Furthermore, the following theorem show that, as time tends to infinity, the system \eqref{eq:obj} itself undergoes approximate convergence to symmetrization.

\begin{theorem}
\label{thm:blen1_in1}
For any $\eta>0, T_1>0$, there exists some $K_c^\ast>0$ such that for any $K_c\geq K_c^\ast$, $\rhb(t)$ generated in \eqref{eq:obj} satisfies
\[
\ba
{
\|\rhb(t)-\mathcal{P}_\ast(\rhb(t))\|_{\rm F}\leq\eta,\quad \forall t\geq T_1}.
\ea
\]
\end{theorem}
The proof of Theorem \ref{thm:blen1_in1} can be seen in Appendix \ref{app:thm4}. The  $K_c^\ast$ in Theorem \ref{thm:blen1_in1} can be given by
\[
K_{c}^\ast =\max\left\{ \frac{1}{T_1\lambda_{\rm m}(\Qs_{\rm d})}\ln\left(\frac{2C}{\eta}\right), \frac{2\|\As_{\rm d}\|}{\eta\lambda_{\rm m}(\Qs_{\rm d})}\right\},
\]
where $C=\|{\rhb}(0)-\mathcal{P}_\ast(\rhb(0))\|_{\rm F}$.

Given that $\mathcal{P}_\ast(\rhb(t))$ is the permutation-invariant projection of $\rhb(t)$, Theorem \ref{thm:blen1_in1} shows that the trajectory $\rhb(t)$ converges within arbitrary precision to a permutation-invariant state as $t \to \infty$, provided the coupling gain $K_c$ is sufficiently large.

In Theorems \ref{cor:rex}-\ref{thm:blen1_in1}, a {\em non-zero} lower bound of time, \( T_1 \), is required to reach the given bound \( \eta \). This is due to the inconsistency between the initial states of each qubit. On the other hand, if the initial state of equation \eqref{eq:obj}, \( \rho(0) \), satisfies \( \rho_j = \rho_k \) for all \( j, k \in \mV \),  then Theorems \ref{cor:rex}-\ref{thm:blen1_in1} hold with \( T_1 = 0 \).


  


\medskip

\begin{figure}[t]
   \centering
 \includegraphics[width=8cm]{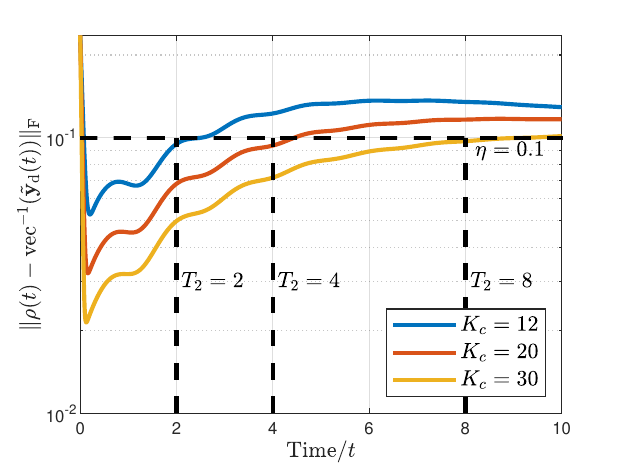}
    \caption{The time evolutions of $\|\rhb(t)-\mathrm{vec^{-1}}(\tilde{\yb}_{\mathrm{d}}(t))\|_{\rm F}$ in \eqref{eq:sim3} with varying diffusive coupling gain $K_c$ for different given $T_2$ and $\eta=0.01$.}
    \label{fig.sim3_2}
\end{figure}

\begin{figure}[t]
   \centering
 \includegraphics[width=8cm]{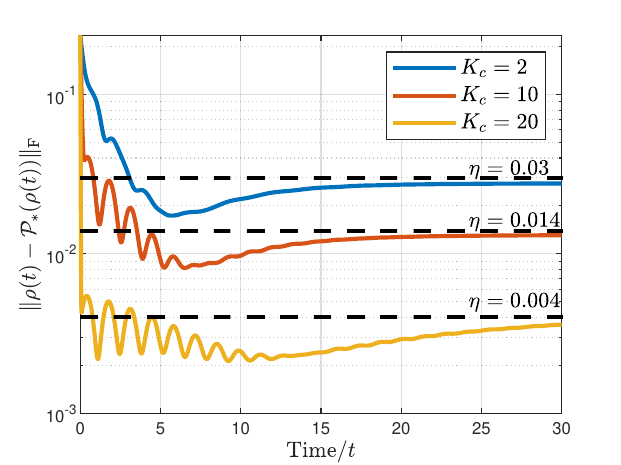}
    \caption{The time evolutions of $\|\rhb(t)-\mathrm{vec^{-1}}(\tilde{\yb}_{\mathrm{d}}(t))\|_{\rm F}$ in \eqref{eq:sim3} with varying diffusive coupling gain $K_c$ for different given $\eta$.}
    \label{fig.sim3_1}
\end{figure}

\noindent{\bf Example 3}. We consider the following Dicke model \cite{dicke} 
\begin{equation}
    \label{eq:sim3}
    \frac{d\rhb}{dt} = -\frac{\imath}{\hbar} [\Hs, \rhb] + \mathcal{D}(\Ls) \rhb + K_c \sum_{\{j,k\} \in \mE} \mathcal{D}(\Us_{jk}) \rhb,
\end{equation}
with the initial state \( \rhb(0) = \frac{1}{6}(|001\rangle\langle001| + 2|010\rangle\langle010| + 3|100\rangle\langle100|) \), where
\[
\ba
&\Hs := \sum_{j=1}^3 \sigma_z^{(j)} + \frac{0.1}{2} \sum_{j=1}^3 (\sigma_+^{(j)} + \sigma_-^{(j)}) + \sum_{j\geq K} \sigma_x^{(j)} \sigma_y^{(k)},\quad
\\&\Ls := \sqrt{0.05} \sum_{j=1}^3 \sigma_-^{(j)} + \sqrt{0.02} \sum_{j=1}^3 \sigma_z^{(j)}.
\ea
\]
Due to the capacitive coupling term \( \sum_{j\geq K} \sigma_x^{(j)} \sigma_y^{(k)} \), the system in \eqref{eq:sim3} is inseparable, making the classical blending approach described in Section~\ref{sec.cla} difficult to apply. However, the blended coherent dynamics, taking the form of \eqref{eq:blen_in}, can still be derived.

The time evolution of the error between the  system \eqref{eq:sim3} and its corresponding blended coherent dynamics  \eqref{eq:blen_in}, measured by \( \|\rhb(t)-\mathrm{vec^{-1}}(\tilde{\yb}_{\mathrm{d}}(t))\|_{\rm F} \), is depicted in Fig.~\ref{fig.sim3_2}. It is evident that, for any fixed \( \eta \) and \( T_2 \), the error remains below \( \eta \) for \( t \leq T_2 \), as long as \( K_c \) is chosen sufficiently large, which verifies Theorem \ref{thm:blen1_in2}.
In addition, the evolution of the distance between $\rho(t)$ and $\mathcal{P}_\ast(\rho(t))$ over time is shown in Fig.~\ref{fig.sim3_1}. This confirms that for any given \( \eta > 0 \), the distance falls below \( \eta \) after a finite time (denoted \( T_1 \) in Theorem \ref{thm:blen1_in1}), provided \( K_c \) is sufficiently large. This verifies Theorem \ref{thm:blen1_in1}.
\hfill$\square$

\section{Conclusion}
\label{sec.con}
We have extended the classical blended dynamics theory to quantum networks and have investigated the collective behavior of nodes under diffusive couplings. We have established the blended reduced-state dynamics for separable Hamiltonian and dissipation cases, showing convergence to equilibrium or periodic trajectories. For non-separable cases, we have introduced the blended coherent dynamics, demonstrating orbital attraction of the network’s density operator. Our theoretical findings have been illustrated through concrete examples, confirming the effectiveness of the proposed frameworks in describing emergent classical and coherent behaviors in open quantum networks.

For future work, we aim to extend our study in several directions. First, we plan to investigate quantum networks with weak couplings between nodes and to establish the corresponding blended dynamics for such scenarios. Second, we intend to generalize our results to discrete-time quantum systems, broadening the applicability of the framework. Last, we hope to apply the theoretical findings presented in this paper to practical implementations, thereby bridging the gap between theory and experimental realizations of quantum network dynamics.

\appendix
\section*{Appendix}

\section{Proof of Theorem \ref{cor:rex}}
\label{app:thm1}

For convenience, we introduce some useful facts related to the Euclidean norm and the Frobenius norm. For operators \(\As\) and \(\Bs\), we have the following relation,
\begin{equation}
    \label{eq:norm}
    \|\As \oplus \Bs\| = \max\{\|\As\|, \|\Bs\|\}.
\end{equation}
For a superoperator \(\Lc\) and a density operator \(\rhb\), the following inequality holds,
\begin{equation}
    \label{eq:lrho}
    \|\Lc \rhb\|_{\rm F} \leq \|\Lc\| \|\rhb\|_{\rm F},
\end{equation}
where \(\|\Lc\|\) denotes the Euclidean norm of the row stacking of \(\Lc\). Additionally, by \cite{swap}, for any density operator \(\rhb\), we have the bound
\begin{equation}
    \label{eq:rhof}
    \|\rhb\|_{\rm F} = \sqrt{\text{tr}(\rhb^2)} \leq 1.
\end{equation}

Now, we proceed with the proof of Theorem \ref{cor:rex}. Consider a general case where \eqref{eq:blen1} has an invariant subspace \(\mathbb{S}\), and we aim to prove
\begin{equation}
    \label{eq:thm1obj}
    \ba
    &E(t) \leq C_1 e^{-\frac{\mu t}{2}}, &&\forall 0 \leq t \leq \frac{2}{\mu} \ln \left( \frac{C_1}{\eta_1} \right), \\
    &E(t) \leq \eta_1, &&\forall t \geq \frac{2}{\mu} \ln \left( \frac{C_1}{\eta_1} \right),
    \ea
\end{equation}
for any given \(\eta_1 > 0\) and some constant \(C_1 > 0\), where \(E(t) := \max_{j \in \mathbb{V}} \left\{\|\rhb_j(t) - \bar{\rhb}(t)\|_{\rm F} + \|\bar{\rhb}(t)\|_{\mathbb{S}}\right\}\), with \(\bar{\rhb}(t) := \frac{1}{n} \sum_{j \in \mathbb{V}} \rhb_j(t)\).

We first analyze the upper bound of \( \max_{j \in \mathbb{V}} \|\rhb_j(t) - \bar{\rhb}(t)\|_{\rm F} \). Define the superoperator \(\Lc_j\) by \( \Lc_j \rhb_j := -\frac{\imath}{\hbar}[\Hs_j, \rhb_j] + \sum_{l \in \mathbb{I}} \gamma_l\mathcal{D}(\Ls_{l,j}) \rhb_j \) for any \(\rhb_j \in \mathbb{C}^{2 \times 2}\). Then, \eqref{eq:partra} can be written as
\begin{equation}
    \label{eq:thm1_1}
    \frac{d \rhb_j}{dt} = \Lc_j \rhb_j + K_c \sum_{\{j,k\} \in \mathbb{E}} a_{jk} (\rhb_k - \rhb_j), \quad \forall j \in \mathbb{V}.
\end{equation}

Since the graph \(\mathrm{G}\) is undirected and connected, its Laplacian \(\Qs\) is positive semi-definite with rank \(n - 1\), as shown in \cite{mag}. According to \cite{iss}, the dynamics of \(\rhb_j(t) - \bar{\rhb}(t)\) in \eqref{eq:thm1_1} is input-to-state stable, yielding the following bound,
\begin{equation}
    \label{eq:pj-bp}
    \ba
    &\sqrt{\sum_{j \in \mathbb{V}} \|\rhb_j(t) - \bar{\rhb}(t)\|_{\rm F}^2} 
    \leq e^{-K_c \lambda_{\rm m}(\Qs) t} \sqrt{\sum_{j \in \mathbb{V}} \|\rhb_j(0) - \bar{\rhb}(0)\|_{\rm F}^2}   + \frac{\sqrt{\sup_{\tau \in [0,t]} \sum_{j \in \mathbb{V}} \|\Lc_j \rhb_j(\tau)\|_{\rm F}^2}}{K_c \lambda_{\rm m}(\Qs)},
    \ea
\end{equation}
where \(\lambda_{\rm m}(\Qs)\) is the minimal nonzero eigenvalue of \(\Qs\).

Noting that
\[
\sup_{\tau \in [0,t]} \sum_{j \in \mathbb{V}} \|\Lc_j \rhb_j(\tau)\|_{\rm F}^2 \leq \sup_{\tau \in [0,t]} \sum_{j \in \mathbb{V}} \|\Lc_j\|^2 \|\rhb_j\|_{\rm F}^2 \leq n \|\Lc\|^2,
\]
where the first inequality follows from \eqref{eq:lrho}, and the last inequality follows from \eqref{eq:rhof} and \eqref{eq:norm} with \(\Lc := \bigoplus_{j \in \mathbb{V}} \mathcal{L}_j\). Substituting this into \eqref{eq:pj-bp}, we obtain the following bound
\begin{equation}
    \label{eq:int1}
    \ba
    \max_{j \in \mathbb{V}} \|\rhb_j(t) - \bar{\rhb}(t)\|_{\rm F} 
    \leq& \sqrt{\sum_{j \in \mathbb{V}} \|\rhb_j(0) - \bar{\rhb}(0)\|_{\rm F}^2} e^{-K_c \lambda_{\rm m}(\Qs) t}  + \frac{n \|\Lc\|}{K_c \lambda_{\rm m}(\Qs)}.
    \ea
\end{equation}

Next, we analyze the upper bound of \( \|\bar{\rhb}(t)\|_{\mathbb{S}} \). By taking the average of the dynamics of \( \rhb_j \), we obtain the time evolution of \( \bar{\rhb} \) as
\begin{equation}
    \label{eq:thm1_2}
    \frac{d\bar{\rhb}}{dt} = \bar{\Lc}\bar{\rhb} + \frac{1}{n} \sum_{j \in \mathbb{V}} \Lc_j(\rhb_j - \bar{\rhb}),
\end{equation}
where we use the fact that \( \bar{\Lc} = \frac{1}{n} \sum_{j \in \mathbb{V}} \Lc_j \). Since the eigenvalues of \( \bar{\Lc} \) have non-positive real parts according to \cite{openquantum}, we assume that the eigenvalues \( \lambda_1, \lambda_2, \dots, \lambda_p \) have zero real parts, and the eigenvalues \( \lambda_{p+1}, \lambda_{p+2}, \dots, \lambda_4 \) have negative real parts. We also assume that \( \hat{\rhb}_1, \hat{\rhb}_2, \hat{\rhb}_3, \hat{\rhb}_4 \) are the corresponding eigenvectors.

According to \cite{sepe}, any \( \rhb \in \mathbb{C}^{2 \times 2} \) can be decomposed in terms of the eigenvectors of \( \bar{\Lc} \) as
\[
\rhb = \sum_{k=1}^4 c_k(\rhb) \hat{\rhb}_k,
\]
for some \( c_k(\rhb) \in \mathbb{C} \), where \( c_k(\rhb) \) is a Lipschitz function with a Lipschitz constant \( L_c > 0 \). Then, we have
\[
e^{\bar{\Lc} t} \rhb = \sum_{k=1}^4 e^{\lambda_k t} c_k(\rhb) \hat{\rhb}_k.
\]
Using this fact, from \eqref{eq:thm1_2}, we have
\[
\ba
    \bar{\rhb}(t) =& \sum_{k=1}^4 e^{\lambda_k t} c_k(\bar{\rhb}(0)) \hat{\rhb}_k  + \frac{1}{n} \sum_{j \in \mathbb{V}} \sum_{k=1}^4 \int_0^t e^{\lambda_k \tau} c_k(\Lc_j(\rhb_j(\tau) - \bar{\rhb}(\tau))) \hat{\rhb}_k \, d\tau,
    \ea
\]
which yields
\[
\ba
    &\bar{\rhb}(t) - \sum_{k=1}^p e^{\lambda_k t} c_k(\bar{\rhb}(0)) \hat{\rhb}_k  + \frac{1}{n} \sum_{j \in \mathbb{V}} \sum_{k=1}^p \int_0^t e^{\lambda_k \tau} c_k(\Lc_j(\rhb_j(\tau) - \bar{\rhb}(\tau))) \hat{\rhb}_k \, d\tau
\\
    = &\sum_{k=p+1}^4 e^{\lambda_k t} c_k(\bar{\rhb}(0)) \hat{\rhb}_k + \frac{1}{n} \sum_{j \in \mathbb{V}} \sum_{k=p+1}^4 \int_0^t e^{\lambda_k \tau} c_k(\Lc_j(\rhb_j(\tau) - \bar{\rhb}(\tau))) \hat{\rhb}_k \, d\tau.
    \ea
\]
Therefore, we have
\begin{equation}
    \label{eq:barph}
\ba
    \|\bar{\rhb}(t)\|_{\mathbb{S}} \leq& L_c \|\bar{\rhb}(0)\|_{\rm F} e^{-\mu t}  + L_c \frac{\max_{j \in \mathbb{V}} \|\Lc_j\|}{n} \int_0^t e^{-\mu \tau} \max_{j \in \mathbb{V}} \|\rhb_j(\tau) - \bar{\rhb}(\tau)\|_{\rm F} \, d\tau\\
    \leq& L_c e^{-\mu t} + \frac{L_c \|\Lc\|}{n} \int_0^t e^{-\mu \tau} \max_{j \in \mathbb{V}} \|\rhb_j(\tau) - \bar{\rhb}(\tau)\|_{\rm F} \, d\tau,
   \ea
\end{equation}
where the first inequality is obtained using \eqref{eq:lrho}, and the last inequality follows from \eqref{eq:rhof} and \eqref{eq:norm}.

Substituting \eqref{eq:int1} into the above result, we obtain
\begin{equation}
    \label{eq:int2}
\ba
    \|\bar{\rhb}(t)\|_{\mathbb{S}} \leq&  \frac{1}{n} L_c \|\Lc\| \int_0^t e^{-\mu \tau} {\Bigg(} \sqrt{\sum_{j \in \mathbb{V}} \|\rhb_j(0) - \bar{\rhb}(0)\|_{\rm F}^2} e^{-K_c \lambda_{\rm m}(\Qs) \tau}  + \frac{n \|\Lc\|}{K_c \lambda_{\rm m}(\Qs)} {\Bigg)} d\tau+L_c e^{-\mu t} 
    \\\leq &L_c e^{-\mu t} + \frac{1}{n} L_c \|\Lc\| \left( \frac{\sqrt{\sum_{j \in \mathbb{V}} \|\rhb_j(0) - \bar{\rhb}(0)\|_{\rm F}^2}}{\mu + K_c \lambda_{\rm m}(\Qs)} \right)  \cdot\left( 1 - e^{-(\mu + K_c \lambda_{\rm m}(\Qs)) t} \right)
    + \frac{L_c \|\Lc\|^2}{\mu K_c \lambda_{\rm m}(\Qs)} (1 - e^{-\mu t}) 
    \\\leq& L_c e^{-\mu t} + \frac{1}{K_c} \frac{L_c \|\Lc\|}{\lambda_{\rm m}(\Qs)} \left( \frac{1}{n} \sqrt{\sum_{j \in \mathbb{V}} \|\rhb_j(0) - \bar{\rhb}(0)\|_{\rm F}^2} + \frac{\|\Lc\|}{\mu} \right).
    \ea
\end{equation}

Using \eqref{eq:int1} and \eqref{eq:int2}, we directly obtain
\[
\ba
    E(t) =& \max_{j \in \mathbb{V}} \left\{ \|\rhb_j(t) - \bar{\rhb}(t)\|_{\rm F} + \|\bar{\rhb}(t)\|_{\mathbb{S}} \right\}
   \leq C_1 e^{-\min\left\{ \mu, K_c \lambda_{\rm m}(\Qs) \right\} t} + \frac{D}{K_c},
    \ea
\]
where \( C_1 := \sqrt{\sum_{j \in \mathbb{V}} \|\rhb_j(0) - \bar{\rhb}(0)\|_{\rm F}^2} + L_c \) and
\(
    D := \frac{n \|\Lc\|}{\lambda_{\rm m}(\Qs)} + \frac{L_c \|\Lc\|}{\lambda_{\rm m}(\Qs)} \left( \frac{1}{n} \sqrt{\sum_{j \in \mathbb{V}} \|\rhb_j(0) - \bar{\rhb}(0)\|_{\rm F}^2} + \frac{\|\Lc\|}{\mu} \right).
\)

Then, by choosing \( K_c \geq K_{c}^\ast(\eta_1) = \max\left\{ \frac{\mu}{\lambda_{\rm m}(\Qs)}, \frac{2D}{\eta_1} \right\} \), we complete the proof of \eqref{eq:thm1obj}. The proof of \eqref{eq:remark} can similarly be completed by setting \( \eta_1 = \eta \) and \( C = C_1 \).

Next, we analyze the case where \eqref{eq:blen1} relaxes to the steady state  \(\rhb_{\rm r}\). Assuming $\|\bar\rhb(t)\|_\mathbb{S}=\|\bar\rhb(t)-\bar\rhb'(t)\|_{\rm F}$ for some $\bar\rhb'(t)\in\mathbb{C}^{2\times2}$, we know that $\bar\rhb'(t)=k(t)\rhb_{\rm r}$ for some $k(t)\in\mathbb{C}$. Then we have\[
|k(t)-1|= \mathrm{tr}(\bar\rhb)-\mathrm{tr}(\bar\rhb')\leq \sqrt{2}\|\bar\rhb-\bar\rhb'\|_{\rm F} \leq \sqrt{2}E(t), 
\]
which gives the following inequality
    \[
    \ba 
     & \max_{j\in\mV}\|\rhb_j-\rhb_{\rm r}\|_{\rm F}\leq E(t) + \|\bar\rhb'-\rhb_{\rm r}\|_{\rm F}
   \leq  E(t)+|k(t)-1|\|\rhb_{\rm r}\|_{\rm F}\leq (1+\sqrt{2})E(t).
    \ea
\]
The last inequality follows from \eqref{eq:rhof}.
Now, using \eqref{eq:thm1obj} and letting $\eta_1=\frac{\eta}{1+\sqrt{2}}$, we conclude that the proof of Theorem \ref{cor:rex} is completed with $C:=(1+\sqrt{2})C_1$ and $K_c\geq K_c^\ast\left(\frac{\eta}{1+\sqrt{2}}\right)$.


\section{Proof of Theorem \ref{thm:blen1_o}}
\label{app:thm2}
We introduce the error state \( \eb(t) := \bar\rhb(t) - \rhb_{\rm b}(t) \). 
Then we have
\[
\ba
\max_{j\in\mV}\|\rhb_j(t)-\rhb_{\rm b}(t)\|_{\rm F}&\leq \max_{j\in\mV}\|\rhb_j(t)-\bar\rhb(t)\|_{\rm F}+\|\eb(t)\|_{\rm F}.
\ea
\]
The time evolution of $\max_{j\in\mV}\|\rhb_j(t)-\bar\rhb(t)\|_{\rm F}$ is given by \eqref{eq:int1}. Next, we will compute the upper bound of $\|\eb(t)\|_{\rm F}$.
From \eqref{eq:thm1_2}, the time evolution of $\eb(t)$ is given by
\[
\ba
\frac{d}{dt} \eb 
&= \bar{\Lc} \eb + \frac{1}{n}\sum_{j\in\mV}\Lc_j(\rhb_j-\bar\rhb).
\ea
\]
Since \( \rhb_{\rm b}(0) = \frac{1}{n} \sum_{j \in \mV} \rhb_j(0)=\bar\rhb(0) \), we conclude that the initial condition is \( \eb(0) = 0 \). Therefore, similar to \eqref{eq:barph}, we obtain
\[
\ba
\eb(t) &= \int_{\tau = 0}^{t} e^{\bar{\Lc}(t - \tau)} \frac{1}{n} \sum_{j\in\mV}\Lc_j(\rhb_j(\tau)-\bar\rhb(\tau)) d\tau, \\
\Rightarrow \|\eb(t)\|_{\rm F} &\leq \frac{\|\Lc\|}{{n}} \int_{\tau = 0}^{t} \|e^{\bar{\Lc}(t - \tau)}\| \max_{j\in\mV}\|\rhb_j(\tau)-\bar\rhb(\tau)\|_{\rm F} \, d\tau.
\ea
\]
Since the eigenvalues of \( \bar\Lc \) have non-positive real parts \cite{openquantum}, based on the results from \cite{matrix}, we know that \( f(t) := M \sum_{j=0}^{m} t^j \), for some \( M > 0 \), provides an upper bound for \( \|e^{\bar{\Lc}(t - \tau)}\| \), where \( m \) denotes the index of the largest block in the Jordan canonical form of \( \Lc \)'s row stacking. Thus, we have
\begin{equation}
    \label{eq:et}
\ba
\eb(t) &\leq f(t) \frac{\|\Lc\|}{{n}} \int_{\tau = 0}^{t} \max_{j\in\mV}\|\rhb_j(\tau)-\bar\rhb(\tau)\|_{\rm F} \, d\tau.
\ea
\end{equation}

Substituting \eqref{eq:int1} into \eqref{eq:et}, we obtain
\begin{equation}
    \label{eq:inter1}
    \ba
    \|\eb(t)\|_{\rm F} \leq &\frac{f(t)\|\Lc\|}{n} \Bigg( \sqrt{\sum_{j\in\mV}\|\rhb_j(0)-\bar{\rhb}(0)\|_{\rm F}^2} 
    \frac{1-e^{K_c\lambda_{\rm m}(\Qs)t}}{K_c\lambda_{\rm m}(\Qs)}  + t \frac{n\|\Lc\|}{K_c\lambda_{\rm m}(\Qs)} \Bigg)\\
\leq &\frac{1}{K_c}\frac{f(t)\|\Lc\|}{\lambda_{\rm m}(\Qs)}\left(\frac{1}{n}\sqrt{\sum_{j\in\mV}\|\rhb_j(0)-\bar{\rhb}(0)\|_{\rm F}^2}+t{\|\Lc\|}\right),
\ea
\end{equation}

With \eqref{eq:inter1} and \eqref{eq:int1}, we have
\[
\ba
\max_{j\in\mV}\|\rhb_j(t)-\rhb_{\rm b}(t)\|_{\rm F}&\leq \max_{j\in\mV}\|\rhb_j(t)-\bar\rhb(t)\|_{\rm F}+\|\eb(t)\|_{\rm F}\leq Ce^{-K_c\lambda_{\rm m}(\Qs)t}+\frac{D(t)}{K_c},
\ea
\]
for $C:=\sqrt{\sum_{j\in\mV}\|\rhb_j(0)-\bar{\rhb}(0)\|_{\rm F}^2}$, $D(t):=\frac{n\|\Lc\|}{\lambda_{\rm m}(\Qs)}+\frac{f(t)\|\Lc\|}{n\lambda_{\rm m}(\Qs)}\sqrt{\sum_{j\in\mV}\|\rhb_j(0)-\bar{\rhb}(0)\|_{\rm F}^2}$.
Next, letting \( K_{c,1}^\ast = \frac{1}{T_1\lambda_{\rm m}(\Qs)}\ln\left(\frac{2C}{\eta}\right)\) and \( K_{c,2}^\ast = \frac{2D(T_2)}{\eta}\), we conclude that 
\[
\ba
\max_{j\in\mV}\|\rhb_j(t)-\rhb_{\rm b}(t)\|_{\rm F}&\leq \eta,\quad\forall t\in[T_1,T_2],
\ea
\]
for \( K_c> \max\{ K_{c,1}^\ast, K_{c,2}^\ast \} \), which completes the proof of Theorem \ref{thm:blen1_o}.

\section{Proof of Theorem \ref{thm:blen1_in2}}
\label{app:thm3}

We define the operator \(\mathsf{S}_{\mathrm{d}}\) whose columns consist of the non-zero eigenvectors of \(\Qs_{\mathrm{d}}\). Then, we have the following relations,
\begin{equation}
    \label{eq:SI_in}
\ba
\Is_{4^n} = \mathsf{S}_{\mathrm{d}} \mathsf{S}_{\mathrm{d}}^{\top} + \Ps_{\mathrm{d}} \Ps_{\mathrm{d}}^{\top}, 
\ea
\end{equation}
\begin{equation}
    \label{eq:lam}
\ba
 \mathsf{S}_{\mathrm{d}}^\top\Qs_{\rm d} \mathsf{S}_{\mathrm{d}}=\Lambda_{\rm d}, 
\ea
\end{equation}
\begin{equation}
    \label{eq:PQ}
\ba
\Ps_{\mathrm{d}} \Qs_{\mathrm{d}} = \mathsf{0},  
\ea
\end{equation}
\begin{equation}
    \label{eq:SIbound_in}
\ba
\|\Ps_{\mathrm{d}}\| &\leq 1, \quad \|\mathsf{S}_{\mathrm{d}}\| \leq 1,
\ea
\end{equation}
where $\Lambda_{\rm d}$ is a diagonal operator whose elements consist of the nonzero eigenvalues of $\Qs_{\rm d}$.

Next, we decompose \(\tilde{\yb}\) by defining \({\nub} := \mathsf{S}_{\mathrm{d}}^{\top} \tilde{\yb}\) and \(\tilde{\xib} := \Ps_{\mathrm{d}}^{\top} \tilde{\yb}\), and introduce  the error state \(\tilde{\eb} := \tilde{\xib} - \yb_{\mathrm{d}}\). Then there holds
\[
\ba
\|\rhb - \mathrm{vec}^{-1}(\tilde{\yb}_{\mathrm{d}})\|_{\rm F} &= \|\tilde{\yb} - \tilde{\yb}_{\mathrm{d}}\| \\
&= \|\tilde{\yb} - \Ps_{\mathrm{d}} \yb_{\mathrm{d}}\| \\
&= \|\mathsf{S}_{\mathrm{d}} {\nub}\| + \|\Ps_{\mathrm{d}} \tilde{\xib} - \Ps_{\mathrm{d}} \yb_{\mathrm{d}}\| \\
&\leq \|{\nub}\| + \|\tilde{\eb}\|,
\ea
\]
where the second equality follows from the fact that \(\tilde{\yb}_{\mathrm{d}} = \Ps_{\mathrm{d}} \yb_{\mathrm{d}}\), the third one is obtained by \(\tilde{\yb} = \mathsf{S}_{\mathrm{d}} {\nub} + \Ps_{\mathrm{d}} \tilde{\xib}\) from \eqref{eq:SI_in}, and the last inequality follows from \eqref{eq:SIbound_in}.

We now compute the time evolution of \(\|\nub(t)\|^2\) as follows,
\[
\begin{aligned}
\frac{d}{dt} \|\nub\|^2
&=  \R(2 \nub^\dagger \mathsf{S}_{\rm d}^{\top} \As_{\rm d} \yb - 2 K_c \nub^\dagger \mathsf{S}_{\rm d}^{\top} \Qs_{\rm d} \yb) \\
&= \R(2 \nub^\dagger \mathsf{S}_{\rm d}^{\top} \As_{\rm d} \yb - 2 K_c \nub^\dagger \mathsf{S}_{\rm d}^{\top} \Qs_{\rm d} (\mathsf{S}_{\rm d} \mathsf{S}_{\rm d}^{\top} + \mathsf{1}_{\rm d} \mathsf{1}_{\rm d}^{\top}) \yb) \\
&= \R(2 \nub^\dagger \mathsf{S}_{\rm d}^{\top} \As_{\rm d} \yb - 2 K_c \nub^\dagger \mathsf{S}_{\rm d}^{\top} \Lambda_{\rm d} \mathsf{S}_{\rm d} \nub)
\\
&\leq 2 \|\nub\| \|\mathsf{S}_{\rm d}\| \|\As_{\rm d}\| \|\yb\| - 2 K_c \lambda_{\rm m}(\Qs_{\rm d}) \| \nub\|^2 \\
&\leq 2 \|\nub\| \|\As_{\rm d}\| - 2 K_c \lambda_{\rm m}(\Qs_{\rm d}) \|\nub\|^2, 
\end{aligned}
\]
with $\lambda_{\rm m}(\Qs)$ being the minimal nonzero eigenvalue of $\Qs$,
where the second equality follows from \eqref{eq:SI_in}, the last equality is derived using \eqref{eq:PQ} and \eqref{eq:lam}, and the last inequality is obtained by \eqref{eq:SIbound_in} and $\|\yb\|=\|\rhb\|_{\rm F}\leq1$ from \eqref{eq:rhof}. 

Thus, we obtain the following bound for $\|\nub(t)\|$ as
\begin{equation}
    \label{eq:vbound}
\ba
\|\nub(t)\| \leq  \|\nub(0)\|   e^{-K_c \lambda_{\rm m}(\Qs_{\rm d}) t} + \frac{\|\As_{\rm d}\|}{K_c \lambda_{\rm m}(\Qs_{\rm d})}.
\ea
\end{equation}

Next, we analyze the time evolution of the upper bound of \(\|\tilde{\eb}(t)\|\). Since \(\tilde{\eb} = \Ps_{\mathrm{d}}^{\top} \tilde{\yb} - \yb_{\mathrm{d}}\), we have
\[
\ba
\frac{d}{dt} \tilde{\eb} &= \Ps_{\mathrm{d}}^{\top} \As_{\mathrm{d}} \tilde{\yb} - K_c \Ps_{\mathrm{d}}^{\top} \Qs_{\mathrm{d}} \tilde{\yb} + \Ps_{\mathrm{d}}^{\top} \As_{\mathrm{d}} \Ps_{\mathrm{d}} \yb_{\mathrm{d}} \\
&= \Ps_{\mathrm{d}}^{\top} \As_{\mathrm{d}} \Ps_{\mathrm{d}} \tilde{\xib} + \Ps_{\mathrm{d}}^{\top} \As_{\mathrm{d}} \mathsf{S}_{\mathrm{d}} {\nub} + \Ps_{\mathrm{d}}^{\top} \As_{\mathrm{d}} \Ps_{\mathrm{d}} \yb_{\mathrm{d}} \\
&= \Ps_{\mathrm{d}}^{\top} \As_{\mathrm{d}} \Ps_{\mathrm{d}} \tilde{\eb} + \Ps_{\mathrm{d}}^{\top} \As_{\mathrm{d}} \mathsf{S}_{\mathrm{d}} {\nub},
\ea
\]
where the first equality follows from \eqref{eq:obj_in} and \eqref{eq:blen_in}, and the last equality follows from \eqref{eq:SI_in}. Noting \(\eb(0) = \mathbf{0}\), we have
\[
\|\eb(t)\| \leq \|\As_{\mathrm{d}}\| \int_{\tau = 0}^{t} \|e^{\Ps_{\mathrm{d}}^{\top} \As_{\mathrm{d}} \Ps_{\mathrm{d}}(t - \tau)}\| \|{\nub}(\tau)\| \, d\tau.
\]
Since \(\Ps_{\mathrm{d}}^{\top} \As_{\mathrm{d}} \Ps_{\mathrm{d}}\) only has eigenvalues with non-positive real parts (as \(\As_{\mathrm{d}}\) does \cite{openquantum}), it follows that \(\|e^{\Ps_{\mathrm{d}}^{\top} \As_{\mathrm{d}} \Ps_{\mathrm{d}} t}\|\) is bounded by \(f_{\mathrm{d}}(t) := M_{\mathrm{d}} \sum_{j=0}^{m_{\rm d}} t^j\) \cite{matrix}, for some \(M_{\mathrm{d}} > 0\) and \(m_{\rm d}\), the index of the largest block in the Jordan canonical form of \(\As_{\mathrm{d}}\). Thus, with \eqref{eq:vbound}, we obtain
\begin{equation}
    \label{eq:ebound}
\ba
\|\tilde{\eb}(t)\| &\leq \frac{f_{\mathrm{d}}(t) \|\As_{\mathrm{d}}\|}{K_c\lambda_{\mathrm{m}}(\Qs_{\mathrm{d}})}\left(\left( 1 - e^{-{K_c \lambda_{\mathrm{m}}(\Qs_{\mathrm{d}}) t}} \right) \|\nub(0)\|   + t{\|\As_{\rm d}\|}\right)
\leq \frac{f_{\mathrm{d}}(t) \|\As_{\mathrm{d}}\|}{K_c\lambda_{\mathrm{m}}(\Qs_{\mathrm{d}})}\left( \|\nub(0)\|   + t{\|\As_{\rm d}\|}\right).
\ea
\end{equation}

With \eqref{eq:vbound} and \eqref{eq:ebound}, we have
\[
\ba
\|\rhb - \mathrm{vec}^{-1}(\tilde{\yb}_{\mathrm{d}})\|_{\rm F} 
&\leq \|{\nub}\| + \|\tilde{\eb}\|\leq Ce^{-K_c \lambda_{\rm m}(\Qs_{\rm d}) t} +\frac{D(t)}{K_c},
\ea
\]
for $D(t):= \frac{\|\As_{\rm d}\|}{\lambda_{\rm m}(\Qs_{\rm d})}+\frac{f_{\mathrm{d}}(t) \|\As_{\mathrm{d}}\|}{\lambda_{\mathrm{m}}(\Qs_{\mathrm{d}})}\left( \|\nub(0)\|   + t{\|\As_{\rm d}\|}\right)$ and $C:=\|\nub(0)\|$. Next, letting \( K_{c,1}^\ast = \frac{1}{T_1\lambda_{\rm m}(\Qs_{\rm d})}\ln\left(\frac{2C}{\eta}\right)\) and \( K_{c,2}^\ast = \frac{2D(T_2)}{\eta}\), we conclude that 
\[
\ba
\|\rhb - \mathrm{vec}^{-1}(\tilde{\yb}_{\mathrm{d}})\|_{\rm F}&\leq \eta,\quad\forall t\in[T_1,T_2],
\ea
\]
for \( K_c\geq \max\{ K_{c,1}^\ast, K_{c,2}^\ast \} \).

Finally, noting that
\begin{equation}
    \label{eq:nubt}
\ba
&\|\nub(t)\|=\|\mathsf{S}_{\mathrm{d}}\nub(t)\|=\|\tilde{\yb}(t)-\mathsf{P}_{\mathrm{d}}\mathsf{P}_{\mathrm{d}}^\top\tilde{\yb}(t)\|
=\|{\rhb}(t)-\mathcal{P}_\ast(\rhb(t))\|_{\rm F},
\ea
\end{equation}
where the first equality follows from $\nub(0)=\mathsf{S}_{\mathrm{d}}^\top\tilde{\yb}(0)\in \col(\mathsf{S}_{\rm d})$, the column space of $\mathsf{S}_{\rm d}$, the second one follows from \eqref{eq:SI_in}, and the last one follows from \eqref{eq:proandinv}. Then we have $C=\|{\rhb}(0)-\mathcal{P}_\ast(\rhb(0))\|_{\rm F}$ and complete the proof of Theorem \ref{thm:blen1_in2}.

\section{Proof of Theorem \ref{thm:blen1_in1}}
\label{app:thm4}

Letting \( K_{c,1}^\ast = \frac{1}{T_1\lambda_{\rm m}(\Qs_{\rm d})}\ln\left(\frac{2\|\nub(0)\|}{\eta}\right)\) and  $K_{c,2}^\ast=\frac{2\|\As_{\rm d}\|}{\eta\lambda_{\rm m}(\Qs_{\rm d})}$, by \eqref{eq:vbound}, we have
\[
\|\nub(t)\|\leq \eta,\quad \forall \, t\geq T_1,
\]
for \( K_c\geq \max\{ K_{c,1}^\ast, K_{c,2}^\ast \} \). 
With \eqref{eq:nubt}, we directly complete the proof of 
Theorem \ref{thm:blen1_in1}.
\bibliographystyle{agsm}
\bibliography{reference}

@article{openquantumnet,
  author = {Delgado-Granados, Luis H. and Krogmeier, Timothy J. and Sager-Smith, LeeAnn M. and Avdic, Irma and Hu, Zixuan and Sajjan, Manas and Abbasi, Maryam and Smart, Scott E. and Narang, Prineha and Kais, Sabre and Schlimgen, Anthony W. and Head-Marsden, Kade and Mazziotti, David A.},
  title = {Quantum Algorithms and Applications for Open Quantum Systems},
  journal = {Chemical Reviews},
  volume = {125},
  number = {4},
  pages = {1823--1839},
  year = {2025},
  month = feb,
  doi = {10.1021/acs.chemrev.4c00428},
  issn = {0009-2665},
  publisher = {American Chemical Society},
}

@book{quantumcontrol1,
  author    = {D. D'Alessandro},
  title     = {Introduction to Quantum Control and Dynamics},
  edition   = {Second},
  year      = {2021},
  publisher = {{Chapman and Hall/CRC}},
  address   = {Boca Raton, FL},
  series    = {Advances in Applied Mathematics},
  isbn      = {978-0-367-50790-9},
}

@article{quantumcontrol2,
  author    = {S. Kuang and D. Dong and I. R. Petersen},
  title     = {Rapid Lyapunov control of finite-dimensional quantum systems},
  journal   = {Automatica},
  year      = {2017},
  volume    = {81},
  pages     = {164--175},
  issn      = {0005-1098},
  doi       = {10.1016/j.automatica.2017.02.041},
}

@article{quantumcom1,
  author    = {T. Makihara and W. Jiang and A. H. Safavi-Naeini},
  title     = {Proposal for superconducting quantum networks using multioctave transduction to lower frequencies},
  journal   = {Phys. Rev. A},
  year      = {2025},
  volume    = {111},
  number    = {1},
  pages     = {012614},
  doi       = {10.1103/PhysRevA.111.012614},
}

@article{quantumcom2,
  title     = {Basics of perfect communication through quantum networks},
  author    = {A. Kay},
  journal   = {Phys. Rev. A},
  volume    = {84},
  issue     = {2},
  pages     = {022337},
  numpages  = {8},
  year      = {2011},
  month     = {Aug},
  publisher = {American Physical Society},
  doi       = {10.1103/PhysRevA.84.022337},
}

@article{quannet2,
  author    = {X. Wang and P. Pemberton-Ross and S. G. Schirmer},
  title     = {Symmetry and subspace controllability for spin networks with a single-node control},
  journal   = {IEEE Transactions on Automatic Control},
  year      = {2012},
  volume    = {57},
  number    = {8},
  pages     = {1945--1956},
  month     = {Aug},
  doi       = {10.1109/TAC.2012.2199165}
}

@article{emission,
  author    = {A. S. Sheremet and M. I. Petrov and I. V. Iorsh and A. V. Poshakinskiy and A. N. Poddubny},
  title     = {Waveguide quantum electrodynamics: Collective radiance and photon-photon correlations},
  journal   = {Reviews of Modern Physics},
  year      = {2023},
  volume    = {95},
  pages     = {015002},
  month     = {Mar},
  doi       = {10.1103/RevModPhys.95.015002}
}

@article{nonH,
  author    = {F. Verstraete and M. M. Wolf and J. I. Cirac},
  title     = {Quantum computation and quantum-state engineering driven by dissipation},
  journal   = {Nature Physics},
  year      = {2009},
  volume    = {5},
  pages     = {633--636},
  month     = {Sep},
  doi       = {10.1038/nphys1342},
}

@misc{mazzarella2015,
      title={Extending robustness and randomization from Consensus to Symmetrization Algorithms}, 
      author={L. Mazzarella and F. Ticozzi and A. Sarlette},
      year={2015},
      eprint={1311.3364},
      archivePrefix={arXiv},
      primaryClass={quant-ph},
}

@inproceedings{Ticozzi2014Symmetrization,
  author    = {F. Ticozzi and L. Mazzarella and A. Sarlette},
  title     = {Symmetrization for Quantum Networks: A Continuous-Time Approach},
  booktitle = {Proceedings of the 21st International Symposium on Mathematical Theory of Networks and Systems (MTNS)},
  year      = {2014},
  month     = jul,
  pages     = {1685--1690},
  address   = {Groningen, The Netherlands}
}

@article{Shi2016Reaching,
  author    = {G. Shi and D. Dong and I. R. Petersen and K. H. Johansson},
  title     = {Reaching a Quantum Consensus: Master Equations That Generate Symmetrization and Synchronization},
  journal   = {IEEE Transactions on Automatic Control},
  year      = {2016},
  volume    = {61},
  number    = {2},
  pages     = {374--387},
  month     = feb,
  doi       = {10.1109/TAC.2015.2440551}
}

@ARTICLE{Shi2017,
  author={G. Shi and S. Fu and I. R. Petersen},
  journal={IEEE Transactions on Automatic Control}, 
  title={Consensus of Quantum Networks With Directed Interactions: Fixed and Switching Structures}, 
  year={2017},
  volume={62},
  number={4},
  pages={2014-2019},
  doi={10.1109/TAC.2016.2590503}}

@inproceedings{Takeuchi2016Distributed,
  author    = {R. Takeuchi and K. Tsumura},
  title     = {Distributed Feedback Control of Quantum Networks},
  booktitle = {Proceedings of the 22nd {IFAC} Workshop},
  series    = {{IFAC}-{P}apers{O}n{L}ine},
  volume    = {49},
  pages     = {309--314},
  year      = {2016},
  doi       = {10.1016/j.ifacol.2016.10.432},
}

@article{Jafarizadeh2016Optimizing,
  author    = {S. Jafarizadeh},
  title     = {Optimizing the Convergence Rate of the Quantum Consensus: A Discrete-Time Model},
  journal   = {Automatica},
  year      = {2016},
  volume    = {73},
  pages     = {237--247},
  doi       = {10.1016/j.automatica.2016.07.002}
}

@article{Shi2017Reaching,
  author    = {G. Shi and B. Li and Z. Miao and P. M. Dower and M. R. James},
  title     = {Reaching Agreement in Quantum Hybrid Networks},
  journal   = {Scientific Reports},
  year      = {2017},
  volume    = {7},
  number    = {1},
  pages     = {13614},
  doi       = {10.1038/s41598-017-13791-5}
}

@misc{marcozzi2021quantumconsensusoverview,
      title={Quantum Consensus: an overview}, 
      author={M. Marcozzi and L. Mostarda},
      year={2021},
      eprint={2101.04192},
      archivePrefix={arXiv},
      primaryClass={quant-ph}, 
note          = {Preprint available at arXiv:2101.04192}
}

@ARTICLE{TAC2016HKim,
  author={J. Kim and J. Yang and H. Shim and J.-S. Kim and J. Seo},
  journal={IEEE Transactions on Automatic Control}, 
  title={Robustness of Synchronization of Heterogeneous Agents by Strong Coupling and a Large Number of Agents}, 
  year={2016},
  volume={61},
  number={10},
  pages={3096-3102},
  doi={10.1109/TAC.2015.2498138}
}

@article{LEE2020108952,
title = {A tool for analysis and synthesis of heterogeneous multi-agent systems under rank-deficient coupling},
journal = {Automatica},
volume = {117},
pages = {108952},
year = {2020},
issn = {0005-1098},
doi = {https://doi.org/10.1016/j.automatica.2020.108952},
author = {J. Lee and H. Shim},
}

@Inbook{Lee2022,
author={J. Lee and H. Shim},
title={Design of Heterogeneous Multi-agent System for Distributed Computation},
bookTitle={Trends in Nonlinear and Adaptive Control: A Tribute to Laurent Praly for his 65th Birthday},
year={2022},
publisher={Springer International Publishing},
address={Cham},
pages={83--108},
doi={10.1007/978-3-030-74628-5_4},
}

@article{synchron,
title = {$H^\infty$ almost state synchronization for homogeneous networks of non-introspective agents: A scale-free protocol design},
journal = {Automatica},
volume = {122},
pages = {109276},
year = {2020},
issn = {0005-1098},
doi = {https://doi.org/10.1016/j.automatica.2020.109276},
author = {Zhenwei Liu and Ali Saberi and Anton A. Stoorvogel and Donya Nojavanzadeh},
}

@article{cluster,
title = {Cluster consensus control of generic linear multi-agent systems under directed topology with acyclic partition},
journal = {Automatica},
volume = {49},
number = {9},
pages = {2898-2905},
year = {2013},
issn = {0005-1098},
doi = {https://doi.org/10.1016/j.automatica.2013.06.017},
author = {Jiahu Qin and Changbin Yu},
}

@book{mag,
  author    = {M. Mesbahi and M. Egerstedt},
  title     = {Graph Theoretic Methods in Multiagent Networks},
  publisher = {Princeton University Press},
  year      = {2010}
}

@article{lind1,
  author  = {G. Lindblad},
  title   = {On the generators of quantum dynamical semigroups},
  journal = {Communications in Mathematical Physics},
  year    = {1976},
  volume  = {48},
  number  = {2},
  pages   = {119--130}
}

@book{lind2,
  author    = {H. Wiseman and G. Milburn},
  title     = {Quantum Measurement and Control},
  publisher = {Cambridge University Press},
  address   = {Cambridge},
  year      = {2010}
}

@book{openquantum,
  author    = {R. Ángel and H. Susana},
  title     = {Open Quantum Systems. An Introduction},
  publisher = {Springer},
  year      = {2012}
}

@article{proj,
  author  = {M. Fabrizio and B. Alberto and B. Nicola and C. Cristiano},
  title   = {Spectral theory of Liouvillians for dissipative phase transitions},
  journal = {Physical Review A},
  year    = {2018},
  volume  = {98},
  number  = {13},
  pages   = {042118}
}

@book{swap,
  author    = {M. Nielsen and I. Chuang},
  title     = {Quantum Computation and Quantum Information},
  edition   = {10th},
  publisher = {Cambridge University Press},
  year      = {2010}
}

@article{swap_operator,
  author  = {L. Mazzarella and A. Sarlette and F. Ticozzi},
  title   = {Consensus for quantum networks: From symmetry to gossip iterations},
  journal = {IEEE Transactions on Automatic Control},
  year    = {2015},
  volume  = {60},
  number  = {1},
  pages   = {158--172}
}

@article{sim1,
  title = {Nonequilibrium magnetic phases in spin lattices with gain and loss},
  author = {Huber, Julian and Kirton, Peter and Rabl, Peter},
  journal = {Phys. Rev. A},
  volume = {102},
  issue = {1},
  pages = {012219},
  numpages = {15},
  year = {2020},
  month = {Jul},
  publisher = {American Physical Society},
  doi = {10.1103/PhysRevA.102.012219},
}

@book{sim2,
  author    = {B. Heinz-Peter and F. Petruccione},
  title     = {The Theory of Open Quantum Systems},
  publisher = {Oxford Academic},
  year      = {2010}
}

@article{dicke,
  author  = {H. Dicke},
  title   = {Coherence in spontaneous radiation processes},
  journal = {Physical Review A},
  year    = {1954},
  volume  = {93},
  pages   = {99--110}
}

@book{matrix,
  author    = {R. Horn and C. Johnson},
  title     = {Matrix Analysis},
  publisher = {Springer},
  year      = {1997}
}

@article{sepe,
  author  = {K. Macieszczak and M. Gu{\c t}{\u a} and I. Lesanovsky and J. Garrahan},
  title   = {Towards a Theory of Metastability in Open Quantum Dynamics},
  journal = {Physical Review Letters},
  year    = {2016},
  volume  = {116},
  number  = {6},
  pages   = {240404}
}

@inproceedings{iss,
  author    = {D. Kingston and W. Ren and R. Beard},
  title     = {Consensus algorithms are input-to-state stable},
  booktitle = {Proceedings of the 2005 American Control Conference},
  year      = {2005},
  volume    = {3},
  pages     = {1686--1690}
}

@article{RevModPhys.82.1209,
  title = {Colloquium: Quantum networks with trapped ions},
  author = {Duan, L.-M. and Monroe, C.},
  journal = {Rev. Mod. Phys.},
  volume = {82},
  issue = {2},
  pages = {1209--1224},
  numpages = {0},
  year = {2010},
  month = {Apr},
  publisher = {American Physical Society},
  doi = {10.1103/RevModPhys.82.1209},
}

@INPROCEEDINGS{WCICA,
  author={Shi, Guodong and Dong, Daoyi and Petersen, Ian R. and Johansson, Karl Henrik},
  booktitle={Proceeding of the 11th World Congress on Intelligent Control and Automation}, 
  title={Consensus of quantum networks with continuous-time markovian dynamics}, 
  year={2014},
  volume={},
  number={},
  pages={307-312},
  doi={10.1109/WCICA.2014.7052732}}

@article{PhysRevResearch.6.013262,
  title = {Numerical simulation of large-scale nonlinear open quantum mechanics},
  author = {Roda-Llordes, M. and Candoli, D. and Grochowski, P. T. and Riera-Campeny, A. and Agrenius, T. and Garc\'{\i}a-Ripoll, J. J. and Gonzalez-Ballestero, C. and Romero-Isart, O.},
  journal = {Phys. Rev. Res.},
  volume = {6},
  issue = {1},
  pages = {013262},
  numpages = {10},
  year = {2024},
  month = {Mar},
  publisher = {American Physical Society},
  doi = {10.1103/PhysRevResearch.6.013262},
}

\end{document}